\documentclass[apj,iop]{emulateapj}

\newcommand{\Ms}{M_\sun}
\newcommand{\Mia}{M_{\rm Ia}}
\newcommand{\Mch}{M_{\rm Ch}}
\newcommand{\Ye}{Y_{\rm e}}
\newcommand{\Mwd}{M_{\rm WD}}
\newcommand{\Mwdi}{M_{\rm WD,0}}
\newcommand{\Mcompi}{M_{\rm comp,0}}
\newcommand{\Mni}{M_{\rm 56Ni}}
\newcommand{\fece}{f_{\rm ECE}}
\newcommand{\fni}{f_{\rm 56Ni}}
\newcommand{\fime}{f_{\rm IME}}
\newcommand{\fco}{f_{\rm CO}}
\newcommand{\rhoc}{\rho_{\rm c}}
\newcommand{\Ek}{E_{\rm kin}}
\newcommand{\En}{E_{\rm nuc}}
\newcommand{\Eb}{E_{\rm bin}}
\newcommand{\bvri}{$BV\!RI$}
\newcommand{\Lbvri}{L_{BV\!RI}}
\newcommand{\uvoir}{$uvoir$}
\newcommand{\Luvoir}{L_{uvoir}}
\newcommand{\Lbi}[1]{L_{BV\!RI,i}^{\rm (#1)}}
\newcommand{\vph}{v_{\rm ph}}
\newcommand{\tb}{t_B}
\newcommand{\tbol}{t_{\rm bol}}

\newcommand{\Emw}{E(\bv)_{\rm MW}}
\newcommand{\Ehost}{E(\bv)_{\rm host}}
\newcommand{\EW}{EW$_{\rm NaID}$}

\renewcommand{\th}{t_{+1/2}}

\shorttitle{Super-Chandrasekhar-Mass LC Models for SN 2009dc}
\shortauthors{Kamiya et al.}

\slugcomment{Accepted for publication in \apj\ on July 18, 2012.}

\begin{document}

\title{Super-Chandrasekhar-Mass Light Curve Models for the Highly Luminous Type I\lowercase{a} Supernova 2009\lowercase{dc}}

\author{Yasuomi Kamiya\altaffilmark{1,2,7}}
\author{Masaomi Tanaka\altaffilmark{3,2}}
\author{Ken'ichi Nomoto\altaffilmark{2,1}}
\author{Sergei I. Blinnikov\altaffilmark{4,5,2}}
\author{Elena I. Sorokina\altaffilmark{5,2}}
\author{Tomoharu Suzuki\altaffilmark{6,1}}

\email{yasuomi.kamiya@ipmu.jp}
\altaffiltext{1}{Department of Astronomy, Graduate School of Science, the University of Tokyo, 7-3-1 Hongo, Bunkyo-ku, Tokyo 113-0033, Japan.}
\altaffiltext{2}{Kavli Institute for the Physics and Mathematics of the Universe, Todai Institutes for Advanced Study, the University of Tokyo, 5-1-5 Kashiwanoha, Kashiwa, Chiba 277-8583, Japan.}
\altaffiltext{3}{National Astronomical Observatory of Japan, 2-21-1 Osawa, Mitaka, Tokyo 181-8588, Japan.}
\altaffiltext{4}{Institute for Theoretical and Experimental Physics, 117218 Moscow, Russia.}
\altaffiltext{5}{Sternberg Astronomical Institute, Lomonosov Moscow State University, 119992 Moscow, Russia.}
\altaffiltext{6}{College of Engineering, Chubu University, 1200 Matsumoto-cho, Kasugai, Aichi 487-8501, Japan.}
\altaffiltext{7}{Research Fellow of the Japan Society for the Promotion of Science.}

\begin{abstract}
Several highly luminous Type Ia supernovae (SNe Ia) have been discovered.
Their high luminosities are difficult to explain with the thermonuclear explosions of the Chandrasekhar-mass white dwarfs (WDs).
In the present study, we estimate the progenitor mass of SN 2009dc, one of the extremely luminous SNe Ia, using the hydrodynamical models as follows.
Explosion models of super-Chandrasekhar-mass (super-Ch-mass) WDs are constructed, and multi-color light curves (LCs) are calculated.
The comparison between our calculations and the observations of SN 2009dc suggests that the exploding WD has a super-Ch mass of 2.2--2.4 $\Ms$, producing 1.2--1.4 $\Ms$ of $^{56}$Ni, if the extinction by its host galaxy is negligible.
If the extinction is significant, the exploding WD is as massive as $\sim$2.8 $\Ms$, and $\sim$1.8 $\Ms$ of $^{56}$Ni is necessary to account for the observations.
Whether the host-galaxy extinction is significant or not, the progenitor WD must have a thick carbon-oxygen layer in the outermost zone (20--30\% of the WD mass), which explains the observed low expansion velocity of the ejecta and the presence of carbon.
Our estimate on the mass of the progenitor WD, especially for the extinction-corrected case, is challenging to the current scenarios of SNe Ia.
Implications on the progenitor scenarios are also discussed.
\end{abstract}

\keywords{supernovae: individual (SN 2009dc) --- radiative transfer --- white dwarfs}

\section{Introduction}

It has been widely accepted that a Type Ia supernova (SN Ia) results from a thermonuclear explosion of a carbon-oxygen (C+O) white dwarf (WD) in a close binary system.
The most likely model is that the explosion is triggered by carbon ignition in the central region of the WD when the WD mass ($\Mwd$) reaches the critical mass \citep[$\Mia$, $\sim$1.38 $\Ms$ for a non-rotating C+O WD; e.g.,][]{Hillebrandt00,Nomoto97,Nomoto00}.
Since $\Mia$ is very close to the Chandrasekhar's limiting mass (Chandrasekhar mass, $\Mch$)\footnote{
Hereafter $\Mch$ denotes the  Chandrasekhar's limiting mass for a {\it non-rotating} C+O WD, i.e., $\Mch=1.46(\Ye/0.5)^2\Ms$, where $\Ye$ is the electron mole fraction \citep[e.g.,][]{Chandrasekhar39}.}%
, the resulting explosions are expected to have similar properties.
Actually, normal SNe Ia are used as standard candles in cosmology \citep{Riess98,Perlmutter99}, after correcting their luminosity dispersion by using the Pskovskii-Phillips relation \citep[e.g.,][]{Pskovskii77,Phillips93}.

Despite their uniformity, several unusual SNe Ia have been found to be much more luminous than normal ones.
They are SN 2003fg \citep{Howell06}, SN 2006gz \citep{Hicken07}, SN 2007if \citep{Scalzo10,Yuan10}, and SN 2009dc \citep{Yamanaka09,Tanaka10,Silverman11,Taubenberger11}.
These SNe Ia all show slow luminosity evolutions \citep[e.g.,][Table 4]{Scalzo10}.
Three of them, except for SN 2003fg, show the clear absorption line of \ion{C}{2} in their early spectra, which are rarely detected for normal SNe Ia \citep[e.g.,][]{Marion06,Tanaka08}.
Such extremely high luminosities require $\gtrsim$1.2 $\Ms$ of radioactive $^{56}$Ni if their explosions are spherically symmetric.
In order to produce such a large amount of $^{56}$Ni, their progenitor C+O WDs are suggested to have super-Chandrasekhar (super-Ch) mass (i.e., $\Mwd>\Mch$), because the exploding WDs should contain more than $\sim$0.3 $\Ms$ of the Si-rich layer and the unburned C+O layer on top of the $^{56}$Ni-rich core \citep{Howell06,Hicken07,Scalzo10,Yuan10,Yamanaka09,Silverman11,Taubenberger11}.
Alternatively, it could also be possible to explain the extremely luminous SNe Ia by asymmetric explosions of Chandrasekhar-mass C+O WDs \citep{Hillebrandt07}.
In this paper, we focus on SN 2009dc and approximate it with a spherically symmetric model, because the spectropolarimetric observations of SN 2009dc suggest that it is a globally spherical explosion \citep{Tanaka10}.

A super-Ch-mass C+O WD model can be formed if it is supported by rapid rotation.
For example, \citet{Hachisu86} constructed two-dimensional models of rapidly rotating WDs.
\citet{Uenishi03} calculated the structure and evolution of two-dimensional C+O WDs that rotate by getting angular momentum form accreting matter.
\citet{Yoon05} investigated the stability of rapidly rotating C+O WDs for a wider parameter range.

Explosions and nucleosynthsis of super-Ch-mass C+O WDs were simulated by \citet{Steinmetz92}, \citet{Pfannes10a}, and \citet{Pfannes10b}.
\citet{Maeda09} studied the bolometric light curves (LCs) of super-Ch-mass WD models.
They constructed the homologously expanding models of super-Ch-mass WDs with parameters of WD mass, $^{56}$Ni mass, abundance distribution, and so on.
\citet{Scalzo10} studied the properties of SN 2007if, assuming a shell-surrounded super-Ch-mass WD model as a result of a WD merger.
They estimated that the ejecta mass is 2.4 $\Ms$ with 1.6 $\Ms$ of $^{56}$Ni.

By applying Arnett's law to the synthesized bolometric LCs, the $^{56}$Ni mass of a SN can be estimated \citep[e.g.,][]{Arnett82}.
\citet{Yamanaka09} and \citet{Silverman11} have suggested that SN 2009dc has $\sim$1.2 $\Ms$ if they neglect the extinction by its host galaxy.
They have also reported that the $^{56}$Ni mass could be as large as $\sim$1.6--1.7 $\Ms$ by taking the extinction into account.
\citet{Taubenberger11} has also analytically estimated that the total mass of SN 2009dc is $\sim$2.8 $\Ms$ and that the ejected $^{56}$Ni mass is $\sim$1.8 $\Ms$.
These extreme values challenge current models and scenarios for SNe Ia.

As described above, all the past works rely on the bolometric LCs, which involves some uncertainties (see Section \ref{LC Calculations}).
In most cases, the analytic method is used to estimate the masses of ejecta and ejected $^{56}$Ni.
To derive more accurate properties of the super-Ch candidates for discussing their progenitor scenarios, more sophisticated models are needed.
In this paper, we calculate multi-color LCs for homologously expanding models of super-Ch-mass WDs for the first time.
Section \ref{Models and Calculations} describes our super-Ch-mass WD models and LC calculations.
In Section \ref{Results}, we compare our results with the photometric and spectroscopic observations of SN 2009dc to estimate the masses of the ejecta and $^{56}$Ni.
Implications of our results are discussed in Section \ref{Discussion}.
We summarize our conclusions in Section \ref{Conclusions}.

\section{Models and Calculations}
\label{Models and Calculations}

In this section, we describe the procedures for constructing our super-Ch-mass WD models and the code for calculating their LCs, to compare with the observations of SN 2009dc.
Since SN 2009dc has a continuum polarization as small as the normal SNe Ia \citep{Tanaka10}, spherical symmetry is assumed.

\subsection{Super-Chandrasekhar-Mass White Dwarf Models}

To construct the homologously expanding models of super-Ch-mass WDs, we apply the approximations similar to those adopted by \citet{Maeda09}.
The models are described by the following parameters.
\begin{itemize}
\item $\Mwd$: total WD mass.

\item $\Mni$ (or $\fni$): mass (or mass fraction) of $^{56}$Ni.
The mass fraction hereafter means the ratio to $\Mwd$; e.g., $\fni=\Mni/\Mwd$.

\item $\fece$: mass fraction of electron-captured elements (ECEs; mostly $^{54}$Fe, $^{56}$Fe, $^{55}$Co, and $^{58}$Ni).
Stable Fe, Co, and Ni are included, all of which are simply assumed to have the same mass fraction.

\item $\fime$: mass fraction of intermediate-mass elements (IMEs).
Si, S, and Ca are included, whose mass fraction ratio is $0.68:0.29:0.03$, 
similar to that in the Chandrasekhar-mass WD model, {\tt W7} \citep{Nomoto84,Thielemann86}.

\item $\fco$: mass fraction of C+O\@.
C and O are contained equally in mass fraction.
\end{itemize}
Since the equation
\begin{equation}
\fece+\fni+\fime+\fco=1\label{sum:f}
\end{equation}
holds, we eliminate $\fime$ in the following discussion.
In total, we have four independent parameters.

The expansion model of a super-Ch-mass WD is constructed as follows.
Firstly, with the above parameters, we calculate the nuclear energy release during the explosion ($\En$) by
\begin{eqnarray}
\frac{\En}{10^{51}\,{\rm erg}}=\left[0.5\fece+0.32\fni+1.24\left(1-\fco\right)\right]\nonumber\\
\times\frac{\Mwd}{\Ms}\label{En}
\end{eqnarray}
\citep[e.g.,][]{Maeda09}.
Then, the binding energy of the WD ($\Eb$) is evaluated by Equations (22) and (32)--(34) described in \citet{Yoon05}.
Here, $\rhoc=3\times10^9$ g cm$^{-3}$ for all models as in \citet{Maeda09}, so that $\Eb$ increases almost linearly with $\Mwd$\@.
For $\Mwd>2.1$ $\Ms$, we extrapolate the formula of $\Eb$ \citep{Jeffery06}.
These $\En$ and $\Eb$ give the kinetic energy of the exploding WD as
\begin{equation}
\Ek=\En-\Eb.\label{Ek}
\end{equation}

How much mass fraction of $^{56}$Ni is synthesized at the deflagration or detonation wave depends mainly on the temperature and thus on the density at the flame front.
For W7, e.g., $^{56}$Ni is synthesized at the flame densities of $\rho\sim\rhoc$--$2\times10^7$ g cm$^{-3}$.
Then the total synthesized $^{56}$Ni mass depends on the mass contained in this density range, and thus on the presupernova density structure of the WD and the flame speed.
The rotating WD is more massive than the non-rotating WD with the same central density \citep{Yoon05}.
For the same central density, therefore, the density profile is shallower for more massive stars, and thus the mass contained in the density range for $^{56}$Ni synthesis is larger; i.e., a more massive WD tends to synthesize more $^{56}$Ni.
Also, for the same central density, faster flame produces more $^{56}$Ni because of less pre-expansion, and the flame speed depends on the WD mass, rotation law, density structure, and possible transition to detonation.
To provide constraints on these physical processes, we calculate the light curve models with various nuclear yields and $\Ek$ for the same central density models.

Next, the structure of the explosion model of a super-Ch-mass WD is obtained by scaling {\tt W7}, the canonical Chandrasekhar-mass WD one, in a self-similar way.
We scale the density ($\rho$) and velocity ($v$) of each radial grid in the model as
\begin{equation}
\rho\propto\Mwd^{5/2}\Ek^{-3/2}
\end{equation}
and
\begin{equation}
v\propto\Mwd^{-1/2}\Ek^{1/2}.\label{scale:v}
\end{equation}

We then consider the distribution of four element groups (ECEs, $^{56}$Ni, IMEs, and C+O).
We assume that $^{56}$Ni in a super-Ch-mass WD model is mixed.
The inner boundary of the mixing region is set to be $v=5000$ km s$^{-1}$ for all the models, because the observed line velocities of \ion{Si}{2} are $>$5000 km s$^{-1}$ \citep{Yamanaka09}.
And we set the outer boundary of the mixing region based on the mass coordinate; $M_r=1.13\left(\Mwd/\Mia\right)$ $\Ms$.
Here the reference mass coordinate ($M_r=1.13$ $\Ms$) is the outer boundary of the $^{56}$Ni distribution in the {\tt W7} model.
This outer boundary corresponds to that of the $^{56}$Ni-produced zone in the {\tt W7} model.
By these definitions, the mixing region is uniquely set when we choose four model parameters. 
Note that, for our models, the velocity at the outer boundary of the mixing region differs among models even with the same $\Mwd$, due to the scaling by Equation (\ref{scale:v}).
This mixing is important to account for the observed velocities of the \ion{Si}{2} line (see Section \ref{Photospheric Velocity}).

\begin{figure*}
\plottwo{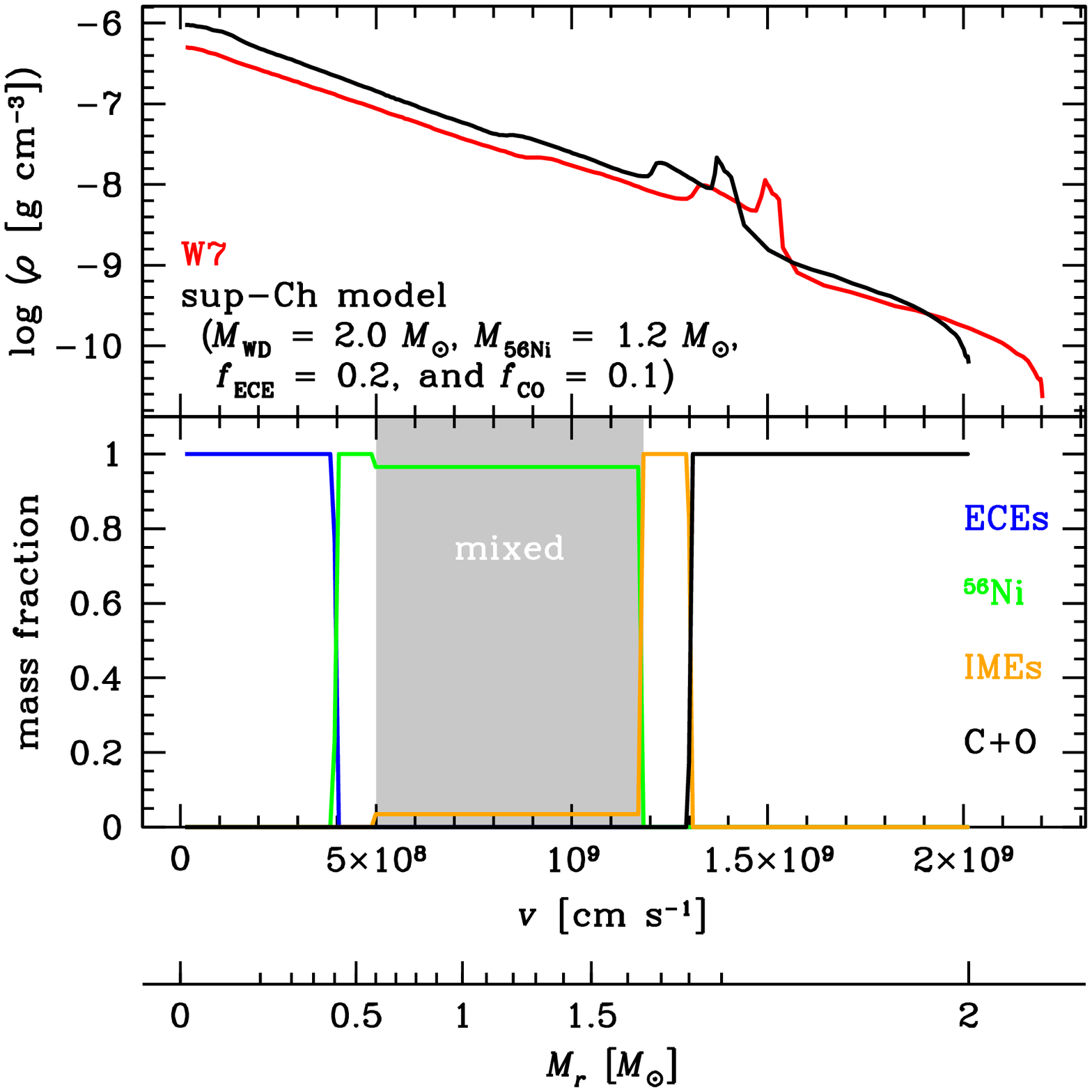}{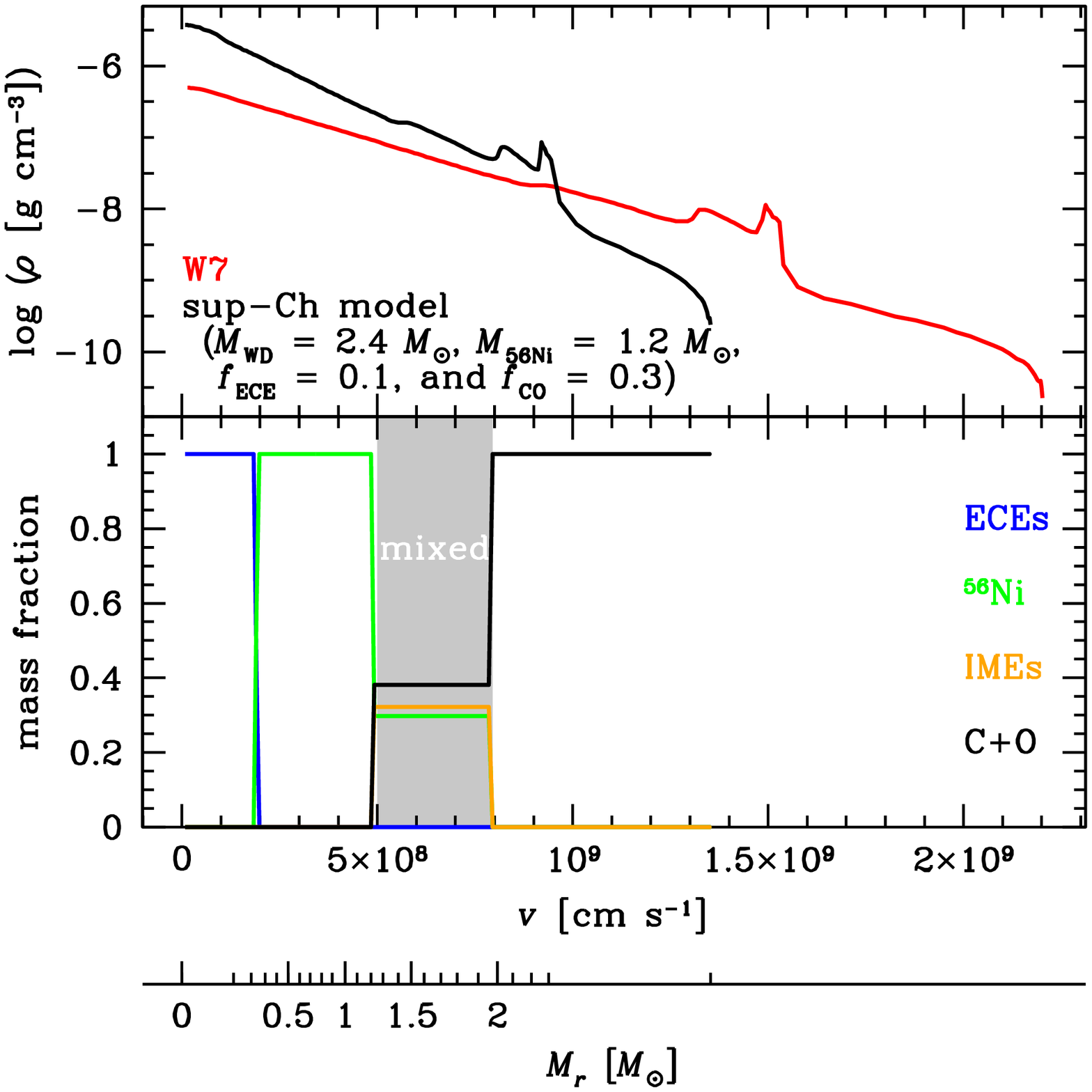}
\caption{Model configurations of two super-Ch-mass WD models as examples.
The parameters of these models are $\Mwd=2.0$ $\Ms$, $\Mni=1.2$ $\Ms$, $\fece=0.2$, and $\fco=0.1$ in the left panels, and $\Mwd=2.4$ $\Ms$, $\Mni=1.2$ $\Ms$, $\fece=0.1$, and $\fco=0.3$ in the right panels.
{\it Top}: density distribution of the model (red) and {\tt W7} (Chandrasekhar-mass WD model; black) at $2\times10^4$ s after the explosion.
{\it Bottom}: abundance distribution of the model.
The lines indicate ECEs (blue), $^{56}$Ni (green), IMEs (orange), and C+O (black).
The mixing region is expressed as the shaded area.\label{model configuration}}
\end{figure*}

To demonstrate the scaling and mixing, we plot the structure and abundance distribution of the two super-Ch-mass WD models in Figure \ref{model configuration}.
The model plotted on the left panels had no IMEs at $5000\lesssim v\lesssim11000$ km s$^{-1}$ before being mixed; the mixing has extended $^{56}$Ni outwards and IMEs inwards.
In some models, C+O are also mixed inwards as in the model right in Figure \ref{model configuration}.

We set the parameter range as follows to compare the models with the observations of SN 2009dc.
\begin{itemize}
\item $\Mwd/\Ms=1.8$, 2.0, $\dots$, 2.8.
\item $\Mni/\Ms=1.2$, 1.4, $\dots$, 1.8, 1.9, 2.0.
\item $\fece=0.1$, 0.2, 0.3.
\item $\fco=0.1$, 0.2, 0.3, 0.4.
\end{itemize}
Here, $\fime$ is obtained by using Equation (\ref{sum:f}).
Any models which have $\fime\leq0$ are not constructed (e.g., models with $\Mwd=2.0$ $\Ms$, $\fece=0.2$, and $\fco\ge0.2$).
We also excluded a model where the inner boundary of the IMEs layer without mixing is located outer than the mixing region (e.g., a model with $\Mwd=2.2$ $\Ms$, $\fece=0.3$, and $\fco=0.1$), because no IMEs extend inwards by mixing.
The parameters and $\Ek$ of our models are listed in Columns 2--5 of Table \ref{table}.

\begin{deluxetable*}{ccccccccc}
\tablecaption{Model Summary \label{table}}
\tablecolumns{9}
\tablehead{
\colhead{Name} & \colhead{$\Mwd$} & \colhead{$\Mni$} & \colhead{$\fece$} & \colhead{$\fco$} & \colhead{$\Ek$} & \multicolumn{2}{c}{WRMSR\tablenotemark{a}} & \colhead{$\vph$ range\tablenotemark{d}}\\
\colhead{} & \colhead{($\Ms$)} & \colhead{($\Ms$)} & \colhead{} & \colhead{} & \colhead{($10^{51}$ erg)} & \colhead{w/o ext.\tablenotemark{b}} & \colhead{w/ ext.\tablenotemark{c}} & \colhead{($10^3$ km s$^{-1}$)}
}
\startdata
{\tt W7} & 1.378\tablenotemark{e} & 0.58\tablenotemark{e} & 0.22\tablenotemark{e} & 0.14\tablenotemark{e} & 1.3\tablenotemark{e} & 2.51 & 3.37 & 13.40--7.50\\
 & 1.8 & 1.2 & 0.1 & 0.1 & 1.49 & 0.76 & 2.18 & 12.58--7.83\\
 & 1.8 & 1.2 & 0.1 & 0.2 & 1.26 & 0.62 & 2.10 & 12.06--8.34\\
 & 2.0 & 1.2 & 0.1 & 0.1 & 1.48 & 0.51 & 2.03 & 12.03--8.16\\
 & 2.0 & 1.2 & 0.1 & 0.2 & 1.23 & 0.43 & 1.97 & 11.37--8.17\\
 & 2.0 & 1.2 & 0.2 & 0.1 & 1.58 & 0.86 & 2.23 & 12.86--8.14\\
 & 2.0 & 1.4 & 0.1 & 0.1 & 1.54 & 0.41 & 1.80 & 11.32--7.52\\
 & 2.2 & 1.2 & 0.1 & 0.1 & 1.46 & 0.46 & 1.98 & 11.53--8.30\\
 & 2.2 & 1.2 & 0.1 & 0.2 & 1.19 & 0.31 & 1.92 & 10.61--7.82\\
{\tt A} & {\bf 2.2} & {\bf 1.2} & {\bf 0.1} & {\bf 0.3} & {\bf 0.92} & {\bf 0.42} & {\bf 1.96} & {\bf 9.34--7.05}\\
 & 2.2 & 1.2 & 0.2 & 0.1 & 1.57 & 0.77 & 2.17 & 12.10--8.45\\
 & 2.2 & 1.2 & 0.2 & 0.2 & 1.30 & 0.62 & 2.12 & 11.24--8.26\\
 & 2.2 & 1.4 & 0.1 & 0.1 & 1.53 & 0.39 & 1.66 & 11.56--7.90\\
 & 2.2 & 1.4 & 0.1 & 0.2 & 1.26 & 0.34 & 1.61 & 10.85--8.09\\
 & 2.4 & 1.2 & 0.1 & 0.1 & 1.45 & 0.32 & 1.87 & 11.10--8.16\\
 & 2.4 & 1.2 & 0.1 & 0.2 & 1.15 & 0.42 & 1.88 & 10.05--7.46\\
{\tt B} & {\bf 2.4} & {\bf 1.2} & {\bf 0.1} & {\bf 0.3} & {\bf 0.85} & {\bf 0.59} & {\bf 1.99} & {\bf 8.59--6.51}\\
 & 2.4 & 1.2 & 0.2 & 0.1 & 1.57 & 0.65 & 2.10 & 11.68--8.57\\
 & 2.4 & 1.2 & 0.2 & 0.2 & 1.27 & 0.51 & 2.07 & 10.74--8.05\\
 & 2.4 & 1.2 & 0.3 & 0.1 & 1.69 & 0.95 & 2.29 & 12.27--8.58\\
 & 2.4 & 1.4 & 0.1 & 0.1 & 1.51 & 0.47 & 1.55 & 11.22--8.05\\
 & 2.4 & 1.4 & 0.1 & 0.2 & 1.22 & 0.42 & 1.56 & 10.25--7.72\\
{\tt C} & {\bf 2.4} & {\bf 1.4} & {\bf 0.1} & {\bf 0.3} & {\bf 0.92} & {\bf 0.52} & {\bf 1.63} & {\bf 8.91--6.99}\\
 & 2.4 & 1.4 & 0.2 & 0.1 & 1.63 & 0.49 & 1.79 & 11.83--8.15\\
 & 2.4 & 1.4 & 0.2 & 0.2 & 1.34 & 0.31 & 1.76 & 10.90--8.19\\
 & 2.4 & 1.6 & 0.1 & 0.1 & 1.58 & 0.95 & 1.21 & 11.50--7.62\\
 & 2.4 & 1.6 & 0.1 & 0.2 & 1.28 & 0.90 & 1.24 & 10.44--7.90\\
 & 2.6 & 1.2 & 0.1 & 0.1 & 1.43 & 0.35 & 1.88 & 10.61--7.84\\
{\tt D} & {\bf 2.6} & {\bf 1.2} & {\bf 0.1} & {\bf 0.2} & {\bf 1.11} & {\bf 0.67} & {\bf 1.86} & {\bf 9.54--7.01}\\
{\tt E} & {\bf 2.6} & {\bf 1.2} & {\bf 0.1} & {\bf 0.3} & {\bf 0.79} & {\bf 0.86} & {\bf 2.22} & {\bf 7.64--5.11}\\
 & 2.6 & 1.2 & 0.1 & 0.4 & 0.47 & 1.61 & 2.74 &  5.92--4.19\\
 & 2.6 & 1.2 & 0.2 & 0.1 & 1.56 & 0.46 & 2.03 & 11.20--8.43\\
 & 2.6 & 1.2 & 0.2 & 0.2 & 1.24 & 0.40 & 1.99 & 10.22--7.69\\
{\tt F} & {\bf 2.6} & {\bf 1.2} & {\bf 0.2} & {\bf 0.3} & {\bf 0.92} & {\bf 0.52} & {\bf 2.07} & {\bf 8.68--6.78}\\
 & 2.6 & 1.4 & 0.1 & 0.1 & 1.50 & 0.49 & 1.52 & 10.71--8.06\\
{\tt G} & {\bf 2.6} & {\bf 1.4} & {\bf 0.1} & {\bf 0.2} & {\bf 1.17} & {\bf 0.56} & {\bf 1.52} & {\bf 9.60--7.35}\\
{\tt H} & {\bf 2.6} & {\bf 1.4} & {\bf 0.1} & {\bf 0.3} & {\bf 0.85} & {\bf 0.73} & {\bf 1.64} & {\bf 8.19--6.42}\\
 & 2.6 & 1.4 & 0.2 & 0.1 & 1.63 & 0.51 & 1.75 & 10.98--8.36\\
 & 2.6 & 1.4 & 0.2 & 0.2 & 1.30 & 0.36 & 1.65 & 10.47--7.99\\
 & 2.6 & 1.6 & 0.1 & 0.1 & 1.56 & 1.05 & 1.15 & 10.89--7.79\\
 & 2.6 & 1.6 & 0.1 & 0.2 & 1.24 & 1.35 & 1.24 & 10.21--7.73\\
 & 2.6 & 1.6 & 0.2 & 0.1 & 1.69 & 0.70 & 1.41 & 11.59--7.82\\
 & 2.6 & 1.8 & 0.1 & 0.1 & 1.62 & 1.41 & 0.91 & 11.01--7.40\\
 & 2.6 & 1.8 & 0.1 & 0.2 & 1.30 & 1.44 & 0.94 & 10.16--7.69\\
 & 2.8 & 1.2 & 0.1 & 0.1 & 1.41 & 0.55 & 1.83 & 10.21--7.50\\
{\tt I} & {\bf 2.8} & {\bf 1.2} & {\bf 0.1} & {\bf 0.2} & {\bf 1.07} & {\bf 0.80} & {\bf 1.89} & {\bf 8.98--6.30}\\
 & 2.8 & 1.2 & 0.1 & 0.3 & 0.72 & 1.48 & 2.70 &  6.40--4.52\\
 & 2.8 & 1.2 & 0.1 & 0.4 & 0.37 & 1.90 & 2.94 &  4.97--3.66\\
 & 2.8 & 1.2 & 0.2 & 0.1 & 1.55 & 0.39 & 1.99 & 10.88--8.23\\
{\tt J} & {\bf 2.8} & {\bf 1.2} & {\bf 0.2} & {\bf 0.2} & {\bf 1.21} & {\bf 0.40} & {\bf 1.97} & {\bf 9.69--7.40}\\
{\tt K} & {\bf 2.8} & {\bf 1.2} & {\bf 0.2} & {\bf 0.3} & {\bf 0.86} & {\bf 0.59} & {\bf 2.06} & {\bf 8.09--6.26}\\
 & 2.8 & 1.2 & 0.3 & 0.1 & 1.69 & 0.73 & 2.17 & 11.51--8.64\\
 & 2.8 & 1.2 & 0.3 & 0.2 & 1.35 & 0.63 & 2.16 & 10.35--7.97\\
 & 2.8 & 1.4 & 0.1 & 0.1 & 1.48 & 0.65 & 1.48 & 10.25--7.85\\
{\tt L} & {\bf 2.8} & {\bf 1.4} & {\bf 0.1} & {\bf 0.2} & {\bf 1.13} & {\bf 0.81} & {\bf 1.52} & {\bf 9.18--6.98}\\
{\tt M} & {\bf 2.8} & {\bf 1.4} & {\bf 0.1} & {\bf 0.3} & {\bf 0.78} & {\bf 0.91} & {\bf 1.78} & {\bf 7.55--5.65}\\
 & 2.8 & 1.4 & 0.2 & 0.1 & 1.62 & 0.33 & 1.66 & 10.98--8.27\\
 & 2.8 & 1.4 & 0.2 & 0.2 & 1.27 & 0.58 & 1.62 & 10.08--7.79\\
 & 2.8 & 1.4 & 0.3 & 0.1 & 1.76 & 0.52 & 1.86 & 11.57--8.46\\
 & 2.8 & 1.6 & 0.1 & 0.1 & 1.54 & 1.13 & 1.13 & 10.42--7.81\\
 & 2.8 & 1.6 & 0.1 & 0.2 & 1.20 & 1.34 & 1.18 &  9.54--7.37\\
 & 2.8 & 1.6 & 0.1 & 0.3 & 0.85 & 1.17 & 1.37 &  7.90--6.35\\
 & 2.8 & 1.6 & 0.2 & 0.1 & 1.68 & 0.74 & 1.35 & 11.15--8.00\\
 & 2.8 & 1.6 & 0.2 & 0.2 & 1.34 & 0.80 & 1.31 & 10.14--7.92\\
 & 2.8 & 1.8 & 0.1 & 0.1 & 1.61 & 1.68 & 0.80 & 10.77--7.61\\
{\tt N} & {\bf 2.8} & {\bf 1.8} & {\bf 0.1} & {\bf 0.2} & {\bf 1.26} & {\bf 1.70} & {\bf 0.88} & {\bf 9.72--7.57}\\
 & 2.8 & 1.9 & 0.1 & 0.1 & 1.64 & 1.78 & 0.74 & 10.68--7.45\\
{\tt O} & {\bf 2.8} & {\bf 1.9} & {\bf 0.1} & {\bf 0.2} & {\bf 1.29} & {\bf 1.85} & {\bf 0.76} & {\bf 9.80--7.60}\\
 & 2.8 & 2.0 & 0.1 & 0.1 & 1.67 & 1.98 & 0.62 & 10.64--7.24\\
\enddata
\tablecomments{The parameters and values of the selected models are bold-faced.}
\tablenotetext{a}{Weighted root-mean-square residual of the \bvri\ LC, calculated by Equation (\ref{WRMSR}).}
\tablenotetext{b}{The extinction by the host galaxy of SN 2009dc is neglected ($\Ehost=0$ mag).}
\tablenotetext{c}{The host-galaxy extinction is corrected ($\Ehost=0.14$ mag).}
\tablenotetext{d}{Maximum and minimum values during $-5\leq\tb\leq25$ days.}
\tablenotetext{e}{\citet{Nomoto84,Thielemann86}.}
\end{deluxetable*}

\subsection{Light Curve Calculations}
\label{LC Calculations}

Multi-color LCs for the constructed models are calculated with STELLA code \citep{Blinnikov98,Blinnikov00a,Blinnikov06}, which solves the one-dimensional equations of radiation hydrodynamics.
STELLA was first developed to calculate LCs of Type II-L supernovae \citep{Blinnikov93}.
Its applications to SNe Ia are described in \citet{Blinnikov00b}, \citet{Blinnikov06}, and \citet{Woosley07}.
In the present study, we use homologously expanding ejecta as input (see Section \ref{Models and Calculations}).
Then the radiation hydrodynamics calculation is performed for each model.
\citet{Maeda09} used the one-dimensional gray radiation transfer code \citep{Iwamoto97,Iwamoto00}, where they assumed simplified opacity for line scatterings.
STELLA, on the other hand, considers 155000 spectral lines in LTE assumption to calculate the opacity with expansion effect more realistically, as well as free-free/bound-free transitions and Thomson scattering.

We note here that the ``bolometric'' luminosity reported from the observations is the \uvoir\ luminosity ($\Luvoir$).
\citet{Yamanaka09} derived  $\Luvoir$ by assuming that (actually observed) integrated \bvri\ luminosity ($\Lbvri$) is 60\% of $\Luvoir$.
This is commonly applied for normal SNe Ia \citep{Wang09a}.
Since it is still unknown whether this assumption is applicable to super-Ch candidates, we directly compare theoretical and observed $\Lbvri$ rather than $\Luvoir$.
This is advantage of our multi-color calculations.
In the calculations, the bolometric, \uvoir, and \bvri\ LCs cover 1--50000 \AA, 1650--23000 \AA, and 3850--8900 \AA, respectively.

\subsection{Extinction}

To compare the calculated and observed LCs, the observational data must be corrected for the extinction.
In order to correct the extinction, three values are needed; the $\bv$ excesses for the Milky Way ($\Emw$) and the host galaxy of SN 2009dc ($\Ehost$), and the ratio of the $V$-band extinction to the \bv\ excess ($R_V$) of the host galaxy ($R_V$ is set to be 3.1 for the Milky Way).
Of these values, $\Ehost$ and $R_V$ of the host galaxy are somewhat difficult to estimate.

The observed \ion{Na}{1} absorption line in the host galaxy suggests that the reddening caused by it is not negligible.
To estimate $\Ehost$, one may use the observed color.
However, the observed \bv\ of SN 2009dc is significantly different from normal SNe Ia, which may suggest that the Lira-Phillips relation \citep{Lira96} should not be applied.
The other method to derive $\Ehost$ is using the equivalent width of the \ion{Na}{1} D absorption line \citep[\EW, e.g.][]{Turatto03}.
But it is also noted that the observed \EW\ has a (relatively) large error \citep{Silverman11,Taubenberger11}.
Also for $R_V$, a non-standard, smaller value ($<$3.1) may be preferred for normal SNe Ia \citep[e.g.][]{Nobili08,Wang09b,Folatelli10,Yasuda10}.
If the host galaxy of SN 2009dc also has $R_V<3.1$, its extinction is overestimated.

To cover most of possible ranges of extinction,  we consider two extreme cases, where the extinction by the host galaxy is negligible ($\Ehost=0$ mag) and significant ($\Ehost=0.14$ mag), respectively.
We set $\Emw=0.71$ mag, and $R_V=3.1$ for the host galaxy (same as for the Milky way).
An extinction law by \citet[][Table 3]{Cardelli89} is applied.

\section{Results}
\label{Results}

\subsection{\bvri\ Light Curves}

\begin{figure*}
\plottwo{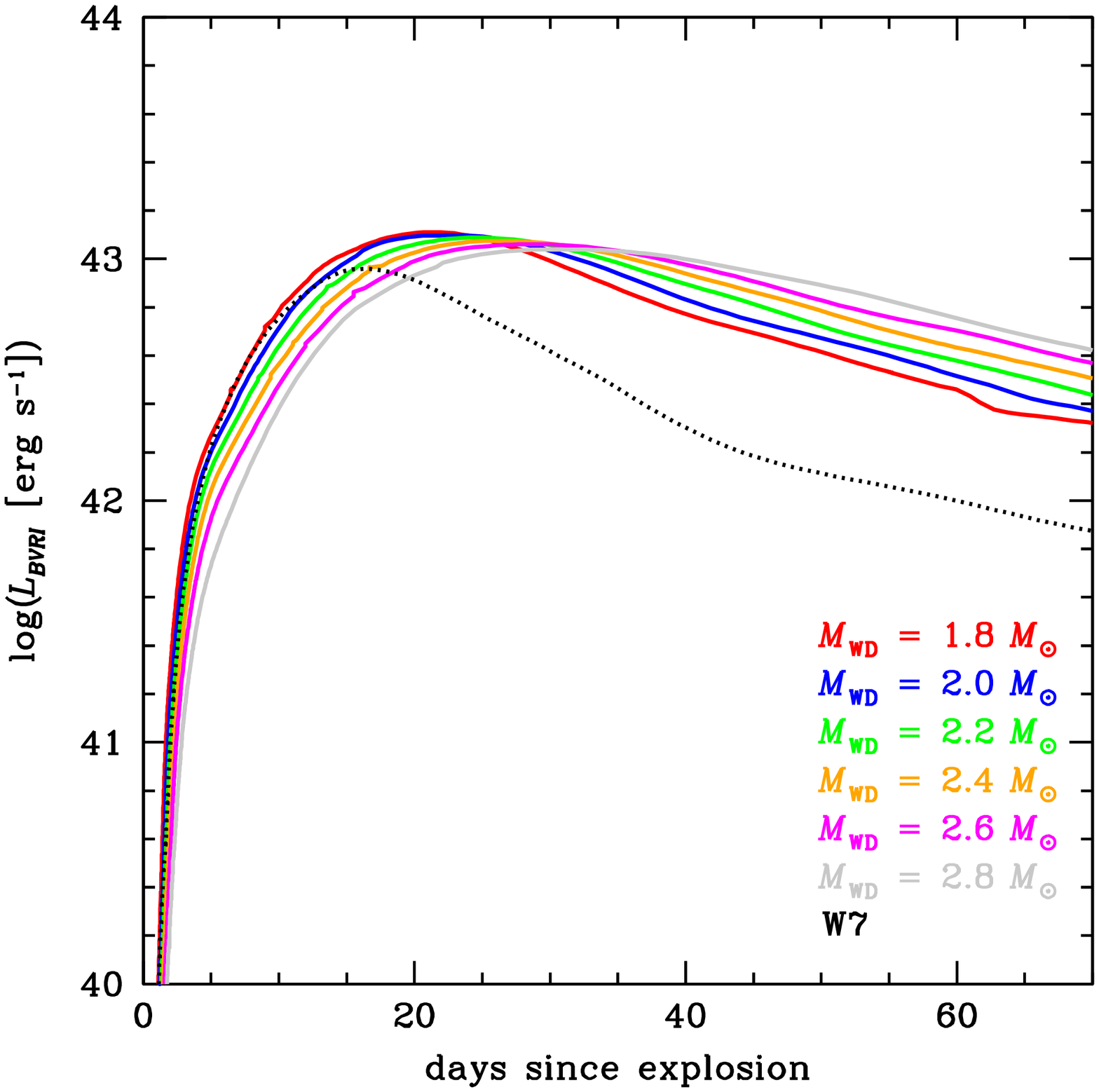}{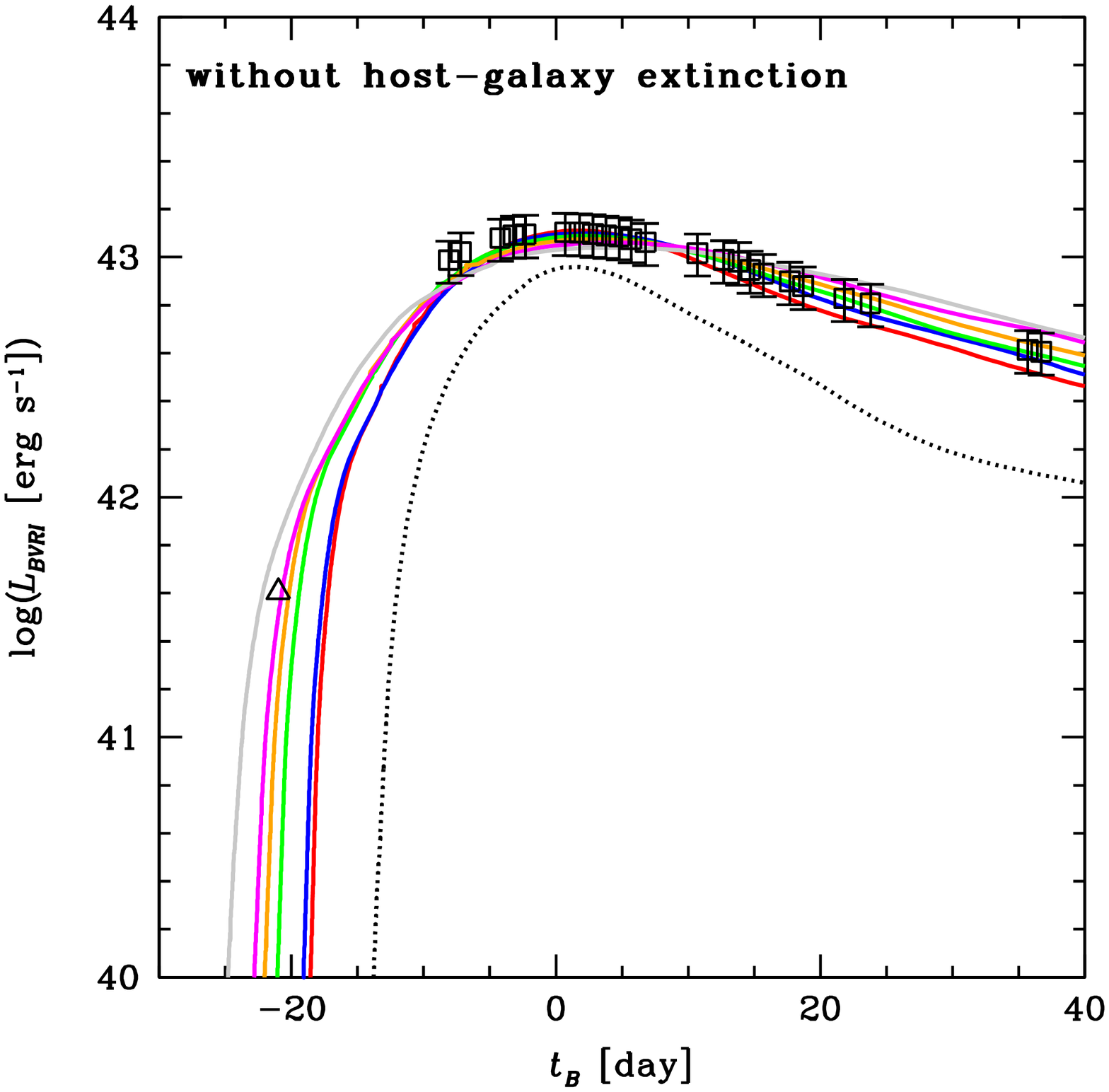}
\caption{{\it Left}: \bvri\ LCs of the super-Ch-mass WD models are plotted (solid lines).
The parameters of the plotted models are $\Mni=1.2$ $\Ms$, $\fece=0.1$, and $\fco=0.2$.
$\Mwd$ of the models are 1.8 $\Ms$ (red), 2.0 $\Ms$ (blue), 2.2 $\Ms$ (green), 2.4 $\Ms$ (orange), 2.6 $\Ms$ (magenta), and 2.8 $\Ms$ (grey).
The \bvri\ LC of {\tt W7} is also plotted (black dotted).
{\it Right}: lines are same as the left plot, but the $x$-axis is now $\tb$.
In addition, the observed LC of SN 2009dc is also shown.
The squares are the observations by \citet{Yamanaka09}, where the error bars are set to be 20\% of $\Lbvri$.
The triangle around $\tb\sim-20$ days indicates the early detection in \citet{Silverman11}, assuming that the $R$-band luminosity is 20\% of the $\Luvoir$ (equivalent to $\Lbvri/3$) and $\Emw=0.07$ mag.\label{bolometric LCs}}
\end{figure*}

The left panel in Figure \ref{bolometric LCs} shows the calculated \bvri\ LCs for the super-Ch-mass WD models with different $\Mwd$.
The other parameters are set to be the same ($\Mni=1.2$ $\Ms$, $\fece=0.1$, and $\fco=0.2$).
A clear relation is seen between the \bvri\ LCs and $\Mwd$ from this panel; a more massive model shows a broader \bvri\ LC\@.

Such a mass dependence is consistent with the relation between the timescale of the bolometric LC ($\tbol$) and $\Mwd$,
\begin{equation}
\tbol\propto\bar{\kappa}^{1/2}\Mwd^{3/4}\Ek^{-1/2}\label{taubol}
\end{equation}
\citep{Arnett82}, where $\bar{\kappa}$ is the opacity averaged in the ejecta (although $\bar{\kappa}$ does differ among models and with time).
Note that $\Ek$ anticorrelates with $\Mwd$ for the models plotted in the left panel of Figure \ref{bolometric LCs} because $\fece$ and $\fco$ are fixed (c.f. Table \ref{table}).

\begin{figure*}
\plottwo{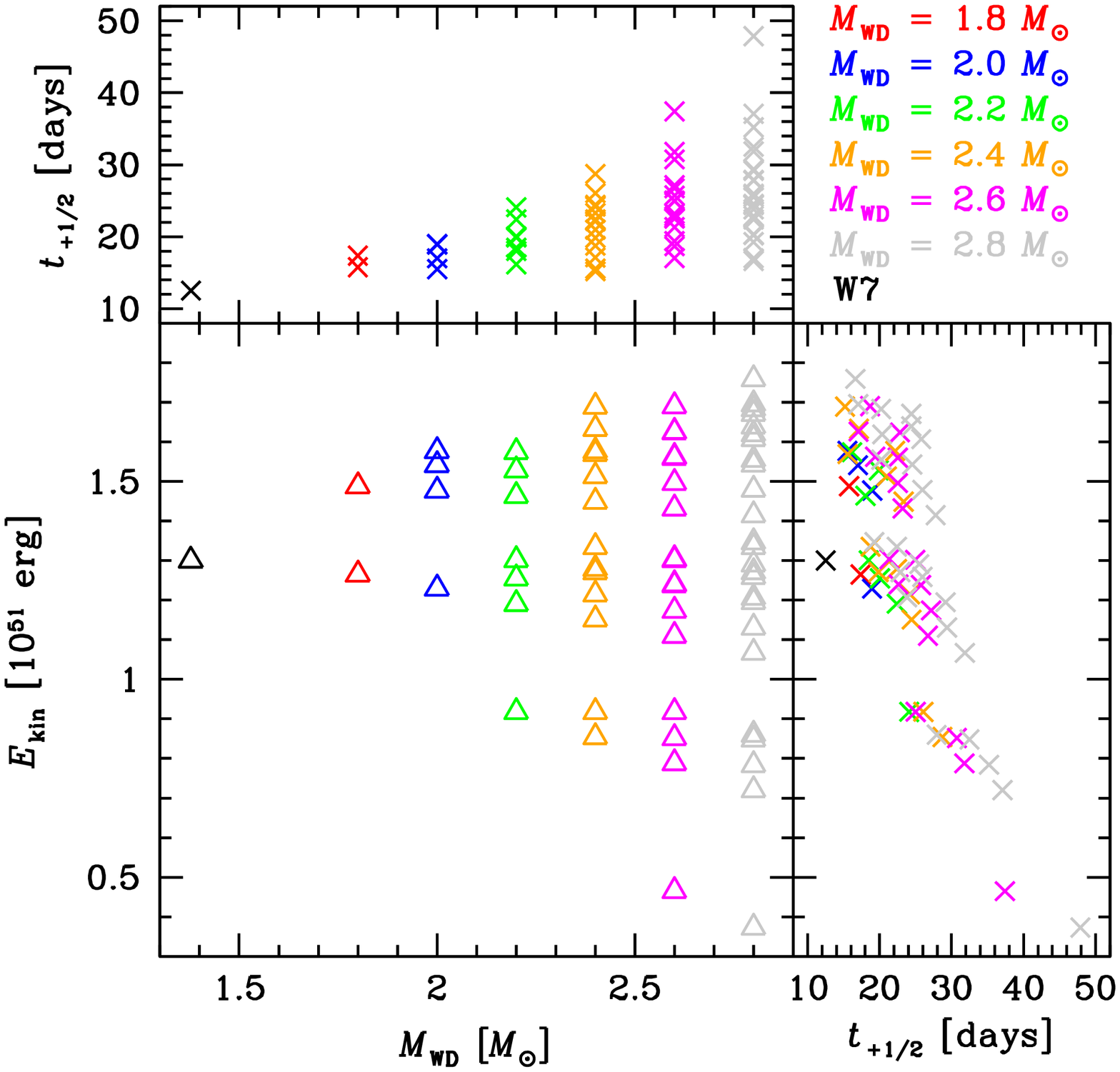}{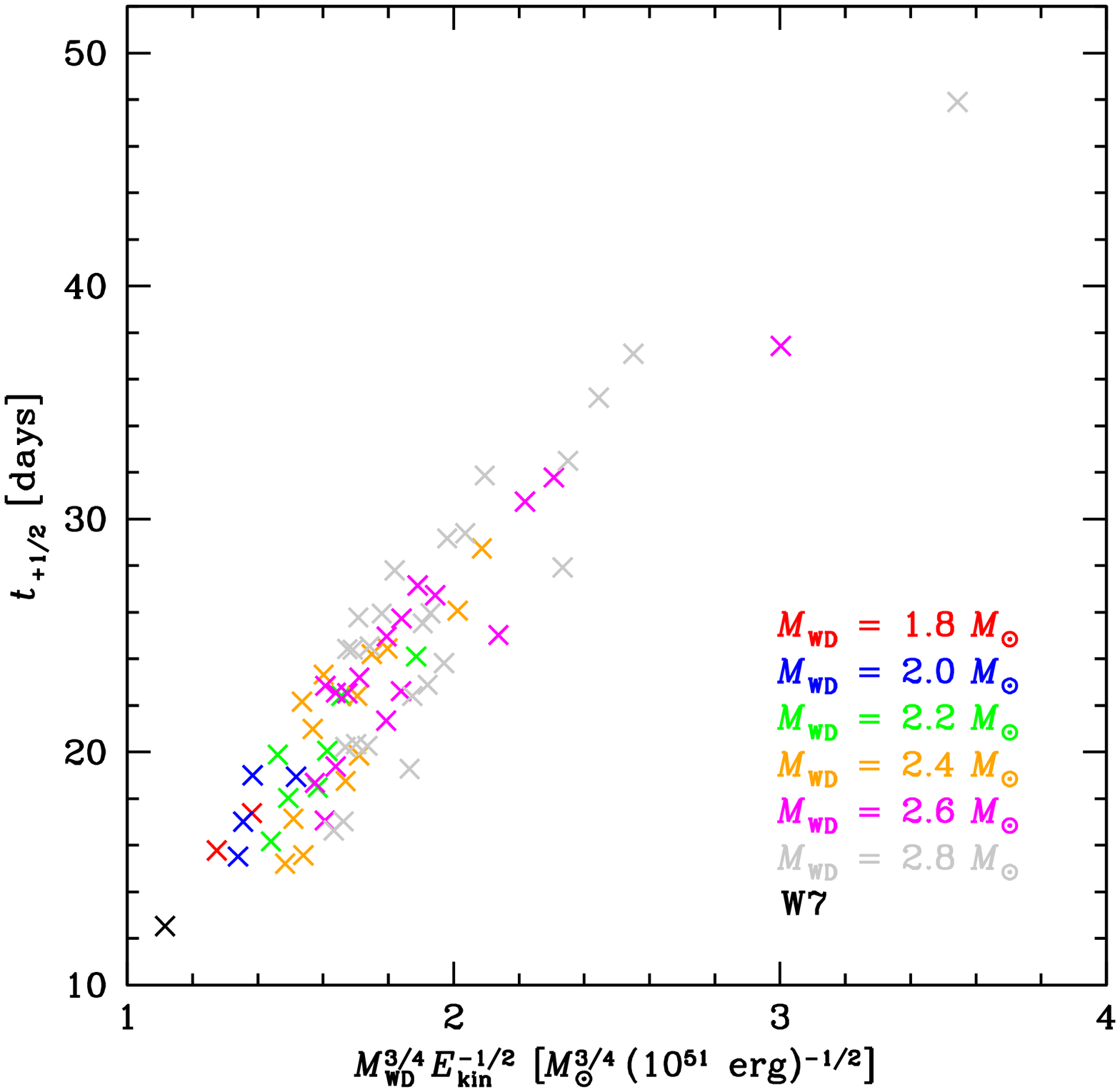}
\caption{{\it Left}: we plot $\Mwd$--$\Ek$ (triangles in the large panel), $\Mwd$--$\th$ (crosses in the small, top panel), and $\Ek$--$\th$ (crosses in the small, right panel) obtained for all the models in this paper.
The colors indicate $\Mwd$ of the models.
{\it Right}: as a function of $\Mwd^{3/4}\Ek^{-1/2}$, we plot $\th$ for all the models.
The crosses are also colored according to $\Mwd$ of the models.\label{f3}}
\end{figure*}

We consider a timescale to see quantitatively if the \bvri\ LCs depends on $\Mwd$ and $\Ek$, analogous to Equation (\ref{taubol}).
For this purpose, we take the declining timescale ($\th$), which is defined as the time for the \bvri\ LC to halve its luminosity after the peak.
In the left panels of Figure \ref{f3}, shown are the $\Mwd$--$\Ek$ (large panel), $\Mwd$--$\th$ (small, top), and $\Ek$--$\th$ (small, right) plots for the models listed in Table \ref{table}.
The \bvri\ LCs of the massive and less energetic models tend to be wider (i.e. larger $\th$), which is expected from the left panel in Figure \ref{bolometric LCs}.
For the dependence of $\th$ on $\Mwd$ and $\Ek$, we plot $\th$ against $\Mwd^{3/4}\Ek^{-1/2}$ in the right panel of Figure \ref{f3}.
An almost linear relation is seen between $\th$ and $\Mwd^{3/4}\Ek^{-1/2}$.
We thus confirm a relation similar to Equation (\ref{taubol}) for the \bvri\ LCs.

The calculated and observed \bvri\ LC are shown on the right panel in Figure \ref{bolometric LCs}, by setting the dates of maximum in the $B$-band of the models and observations at the same day as $\tb=0$ day.
We plot the observational data provided by \citet{Yamanaka09}, neglecting the extinction by the host galaxy (open squares).
The open triangle corresponds to the early $R$-band detection by \citet{Silverman11}, where we simply assume that the $R$-band luminosity is equivalent to 20\% of $\Luvoir$, i.e., $1/3$ of $\Lbvri$, assuming $\Luvoir=\Lbvri/0.6$ \citep[cf.][Figure 24]{Wang09a}.
The \bvri\ LCs of the super-Ch-mass WD models on the right panel are as luminous as the observations around the maximum, while some have relatively broader \bvri\ LCs or shorter rising time than the observations.
Especially, the two less massive models ($\Mwd=1.8$--2.0 $\Ms$) are not preferred because they are too faint at $\tb\sim-20$ days.

In order to find the well-fitted models, we calculate the weighted root-mean-square residual (WRMSR) of the \bvri\ LC,
\begin{equation}
{\rm WRMSR}=\sqrt{\frac{1}{N}\sum_{i=1}^{N}\left(\frac{\Lbi{calc}-\Lbi{obs}}{\delta\Lbi{obs}}\right)^2}\label{WRMSR}
\end{equation}
for each model.
We use the observational data for $-10<\tb<40$ days taken from \citet{Yamanaka09}, where the total number of the observations, $N$, is 23 for this time range.
The early detection by \citet{Silverman11} is excluded in calculating the WRMSR.
Here, $\Lbi{obs}$ and $\Lbi{calc}$ respectively denote $\Lbvri$ of the observation and calculation obtained at the $i$-th epoch.
$\delta\Lbi{obs}$ is the observational error of $\Lbi{obs}$, which is set to 20\% of $\Lbi{obs}$.
\citet{Maeda09} used the decline rate of the bolometric LC after maximum for comparison between their models and observations.
We use the above WRMSR instead of the decline rate, so that brightness can also be taken into account.
The WRMSR is calculated with the observational error, and shows which model fits relatively to the observed \bvri\ LC.

We list the WRMSR of the models for the two extinction cases in Columns 7 and 8 of Table \ref{table}.
Good agreement is found between the calculated and observed \bvri\ LCs for whole range of $\Mwd$ in this study, if the extinction by the host galaxy is not considered.
In the case where the extinction is corrected, most models have larger WRMSR (due to smaller $\Lbvri$), while some models with large $\Mwd$ and $\Mni$ are better fitted.

\subsection{Photospheric Velocity}
\label{Photospheric Velocity}
Next, we compare the photospheric velocity ($\vph$) of the models and the observed line velocity of \ion{Si}{2}\@.
In the last column of Table \ref{table}, listed are the $\vph$ ranges of our models at $-5\leq\tb\leq25$ days.
This period covers the whole phases of the spectroscopic observations by \citet{Yamanaka09}.
We estimate the position of the photosphere (hence, $\vph$) in the model 
from the optical depth at the $R$-band, where 
the rest-frame wavelength of the observed \ion{Si}{2} line (6355 \AA) is located.

The velocity of the observed \ion{Si}{2} line is reported as $<$9000 km s$^{-1}$ at the period \citep{Yamanaka09}.
Since the \ion{Si}{2} absorption line is formed by the Si above the photosphere, the calculated $\vph$ should be smaller than the observed \ion{Si}{2} line velocity \citep[for further details, see][Figure 1]{Tanaka11}.
For the models with $\Mwd=1.8$ $\Ms$ and 2.0 $\Ms$, however, the calculated $\vph$ are $>$10000 km s$^{-1}$, much larger than the observed line velocity of \ion{Si}{2}.
Thus, these less massive models are far from consistent with SN 2009dc.
On the other hand, many models with $\Mwd\geq2.2$ $\Ms$ have $\vph<10000$ km s$^{-1}$ during $-5\leq\tb\leq25$ days, which are somewhat compatible with observations.
These models with small $\vph$ commonly have such a large C+O mass fraction as $\fco=0.2$--0.4.
From Equation (\ref{En}), $\En$ and thus $\Ek$ are affected mainly by $\fco$ rather than $\fece$\@.
A model with larger $\fco$ has smaller $\En$, thus smaller $\Ek$ and $\vph$, being preferred for SN 2009dc.

\subsection{Plausible Models for SN 2009dc}

\begin{figure*}
\plottwo{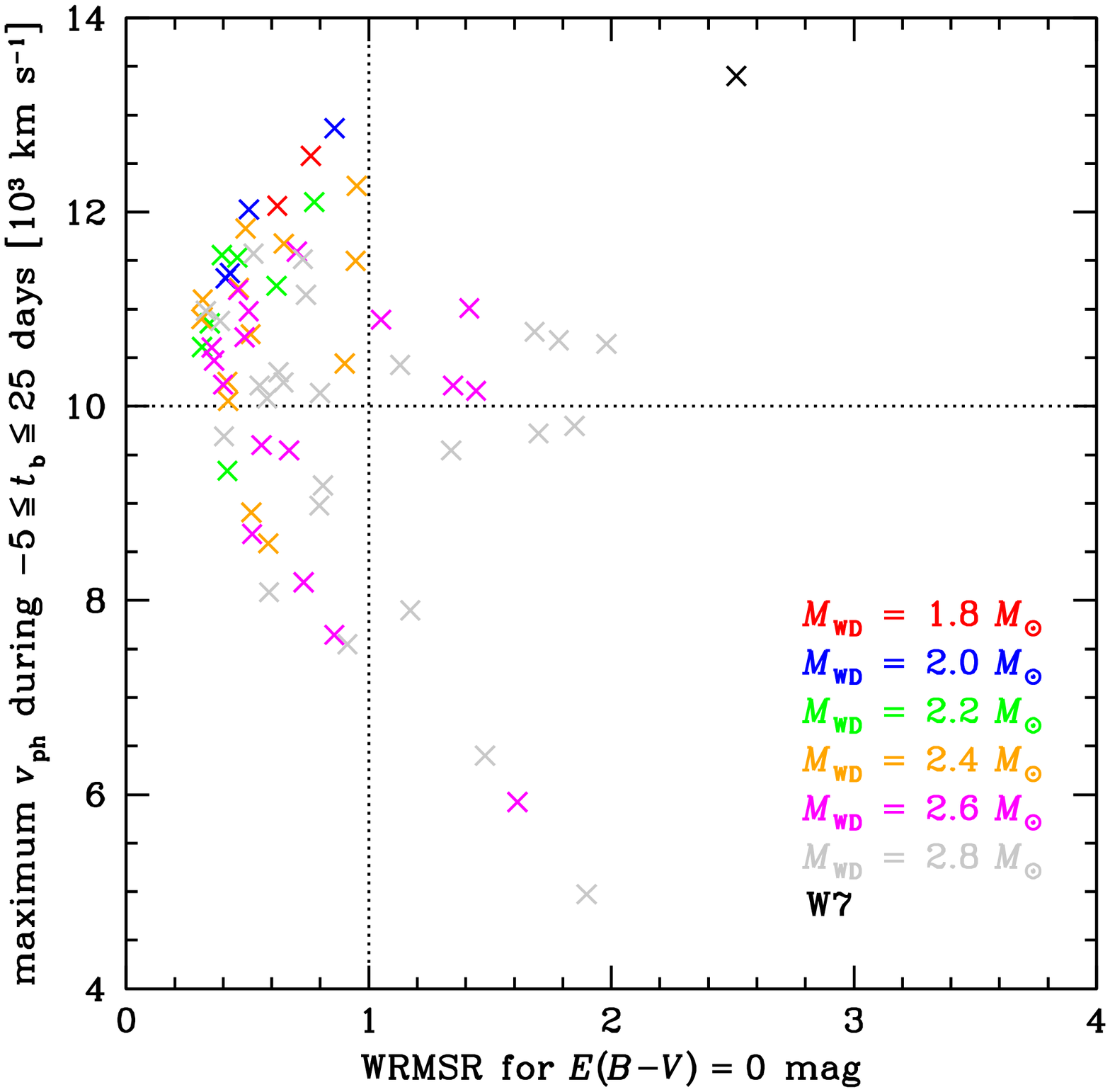}{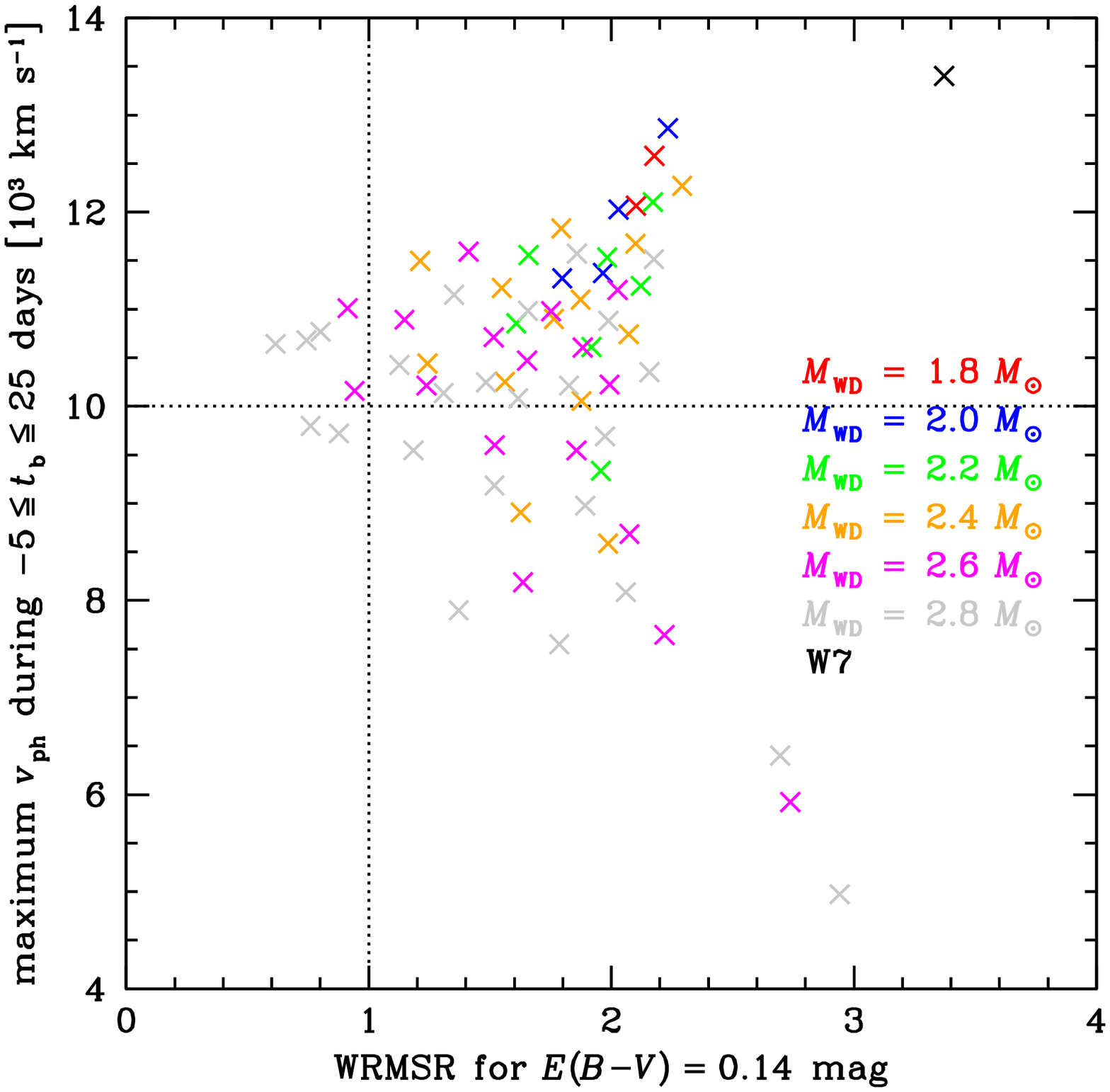}
\caption{Plots for the WRMSR calculated by Equation (\ref{WRMSR}) and the maximum $\vph$ during $-5\leq\tb\leq25$ days of the models.
The crosses with the same color correspond to the models with the same $\Mwd$.
The two dotted lines indicate the criteria for the model selection (see the text).
{\it Left}: $\Ehost=0$ mag is assumed; i.e. the extinction by the host galaxy is not considered in calculating the WRMSR.
{\it Right}: the significant extinction ($\Ehost=0.14$ mag) is assumed.\label{f4}}
\end{figure*}

In order to find the most plausible model for SN 2009dc, we first select models based on \bvri\ LC and $\vph$ discussed above.
We use criteria of WRMSR $\lesssim1$ and $\vph<10000$ km s$^{-1}$ during the $-5\leq\tb\leq25$ days.
By these criteria, models with a label {\tt A}--{\tt O} in Column 1 of Table \ref{table} are selected.
Among them, models {\tt A}--{\tt M} are selected for the case where the host-galaxy extinction is neglected; only models {\tt N} and {\tt O} for the significant extinction.
In Figure \ref{f4}, we plot WRMSR and maximum $\vph$ during the period, where the dotted lines indicates the criteria.

\begin{figure*}
\plottwo{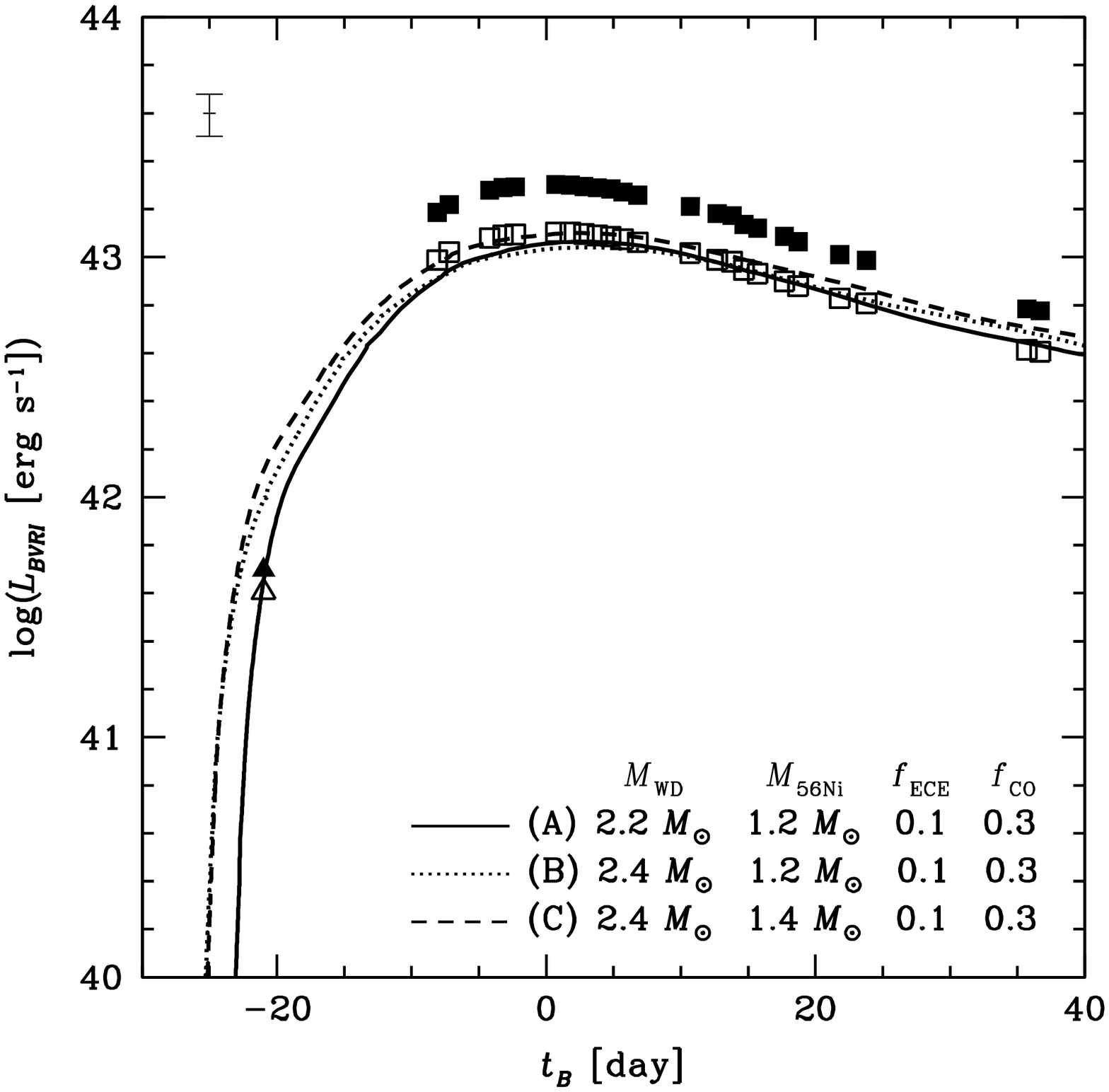}{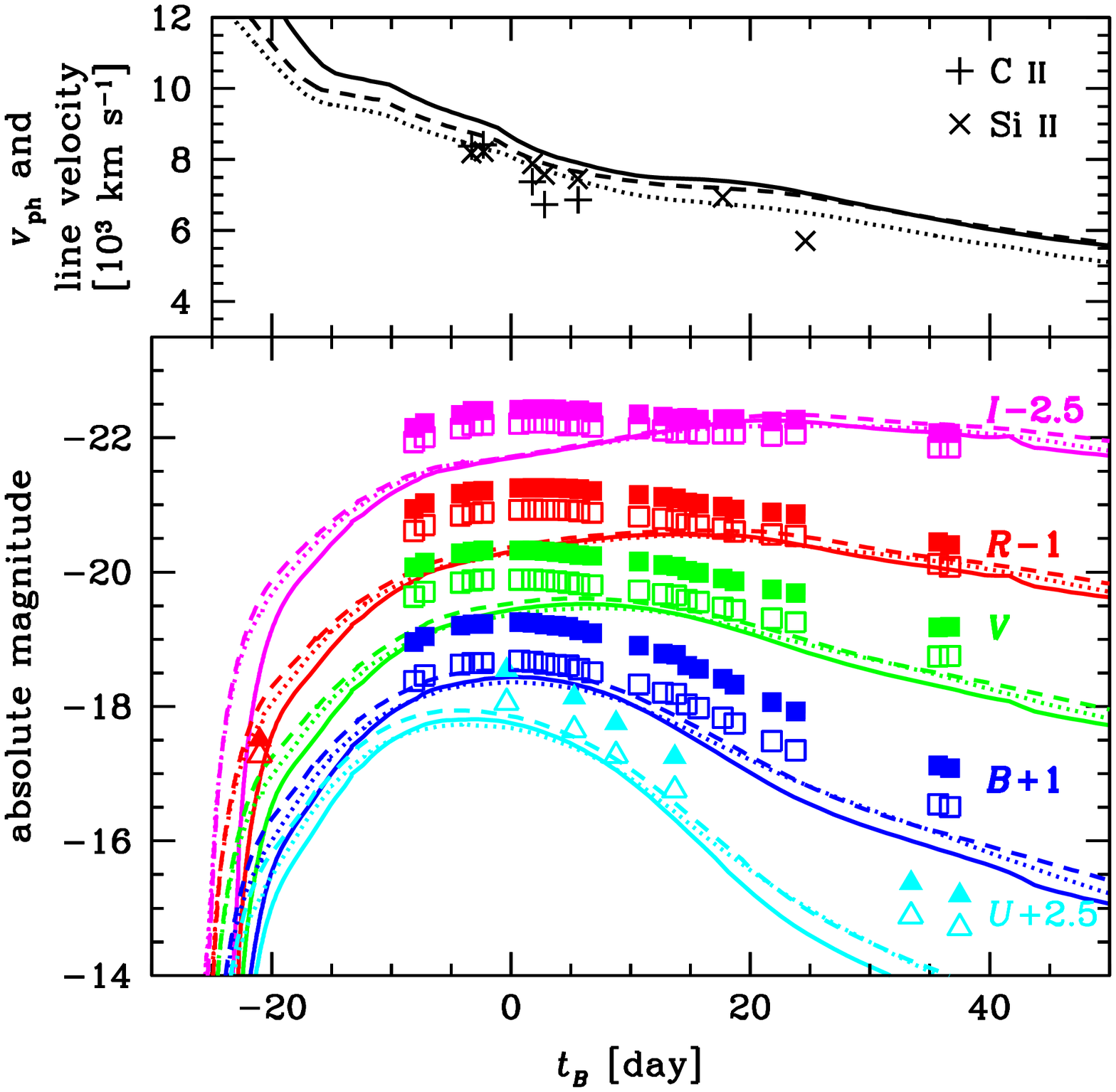}
\caption{\bvri\ and monochromatic LCs, and $\vph$ for models {\tt A}, {\tt B}, and {\tt C}.
{\it Left}: same as the right panel of Figure \ref{bolometric LCs}.
For the open points, we consider the extinction by the host galaxy; for the solid ones, we neglect it.
The point at the left-top corner indicates the error bars for squares.
{\it Right top}: $\vph$ with the line velocity of \ion{C}{2} (pluses) and \ion{Si}{2} (crosses) observed by \citet{Yamanaka09}.
{\it Right bottom}: same as the left panel, but multi-band LCs are plotted.
The observation data in $U$-band (cyan triangles) and the early detection in $R$-band (red triangles around $\tb\sim-20$ days) are taken from \citet{Silverman11}.
Note that we assume $\Emw=0.07$ mag and $\Ehost=0$ (open triangles), 0.10 mag (filled triangles).\label{best-fitted model}}
\end{figure*}

We further analyze multi-color LCs of these models.
Figures \ref{best-fitted model}--\ref{best-fitted model2} show the \bvri\ (left), $\vph$ (right top) monochromatic LCs (right bottom) of these models.
In these figures, the \bvri, $B$-, $V$-, $R$-, and $I$-band LCs by \citet{Yamanaka09} and the early detection and $U$-band LC by \citet{Silverman11} are plotted as squares and triangles, respectively.
Filled and open symbols show the observed LCs with and without the extinction correction.

\begin{figure*}
\plottwo{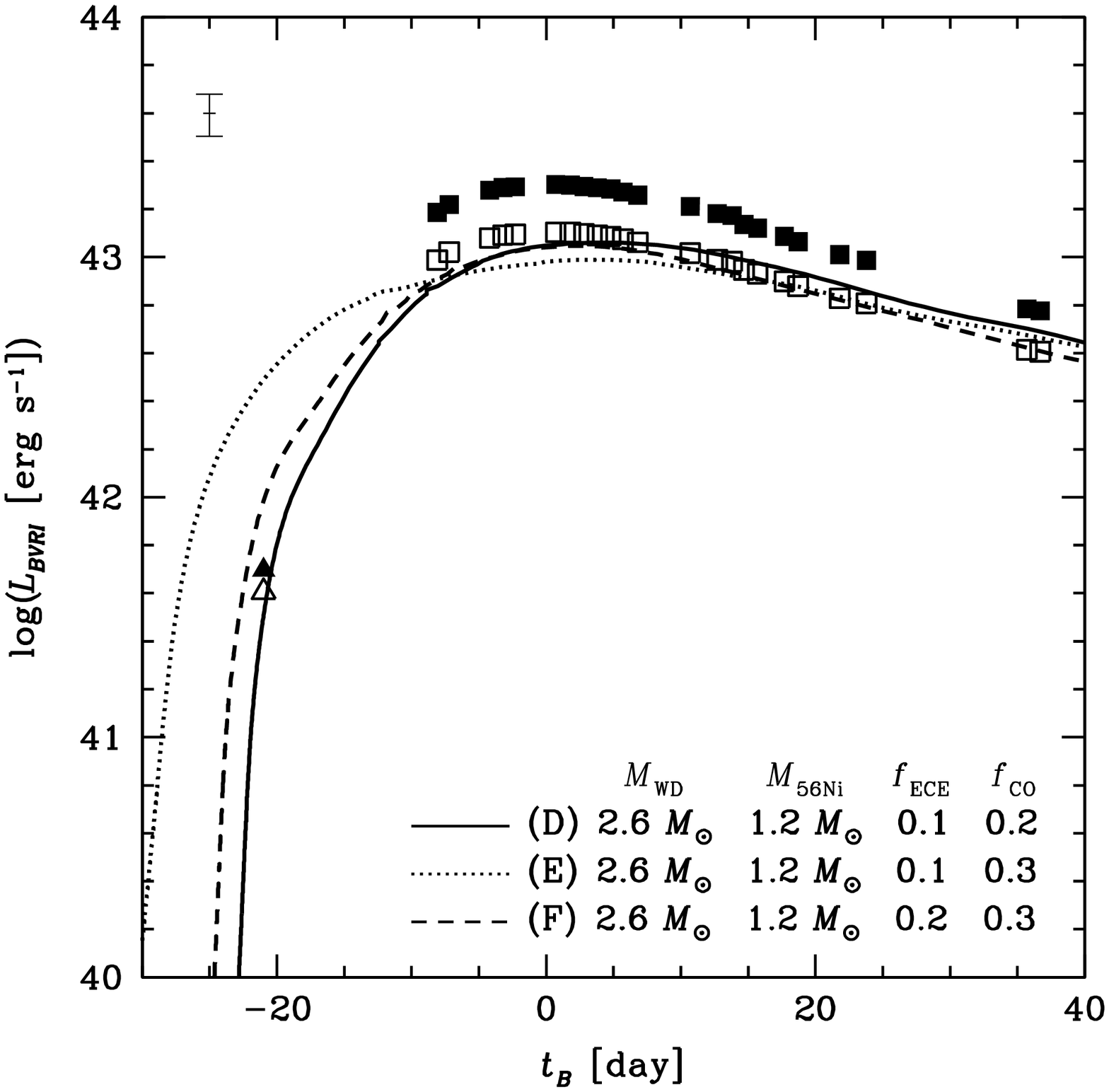}{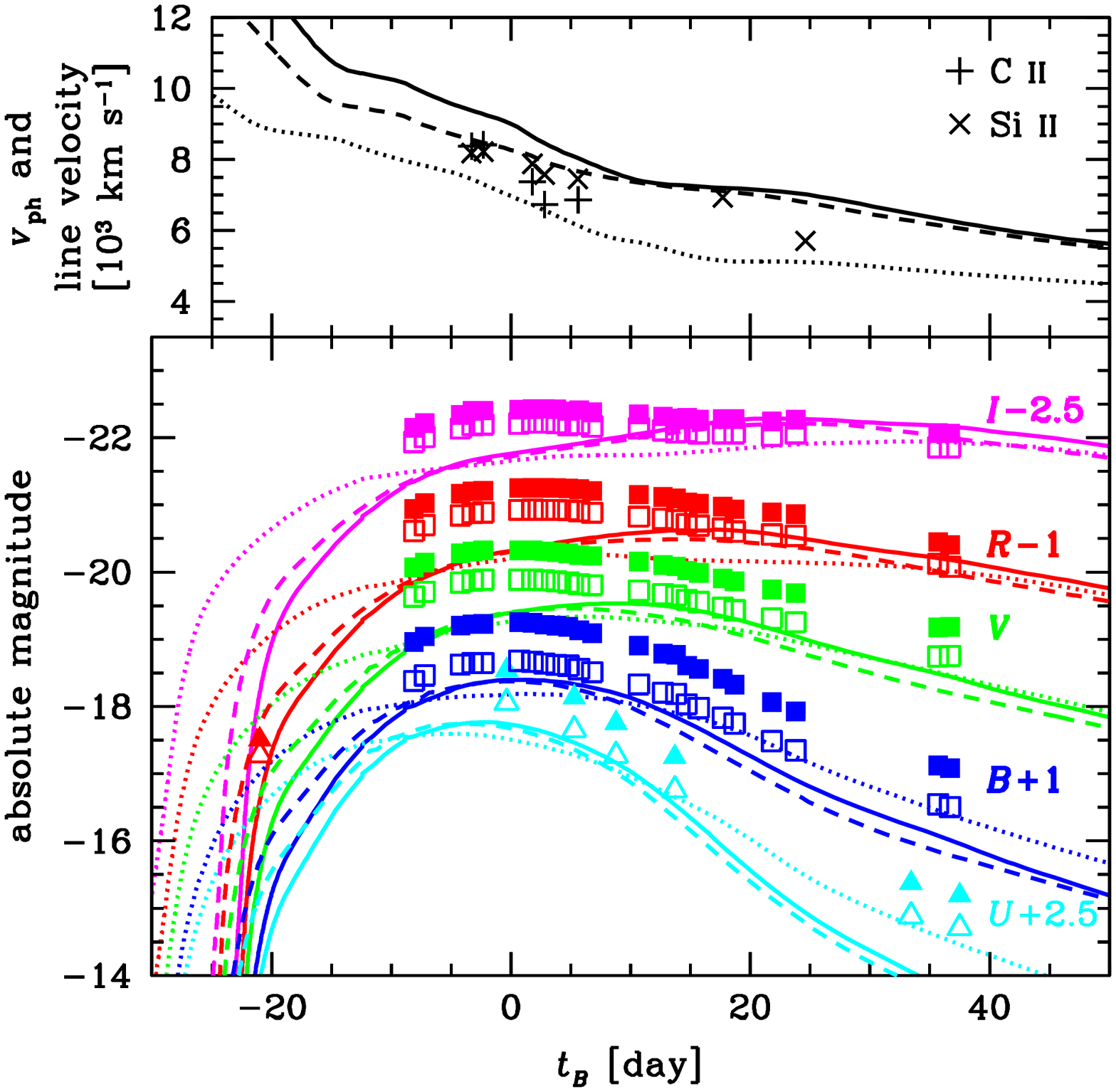}
\caption{Same as Figure \ref{best-fitted model}, but for models {\tt D}, {\tt E}, and {\tt F}.}
\end{figure*}

\begin{figure*}
\plottwo{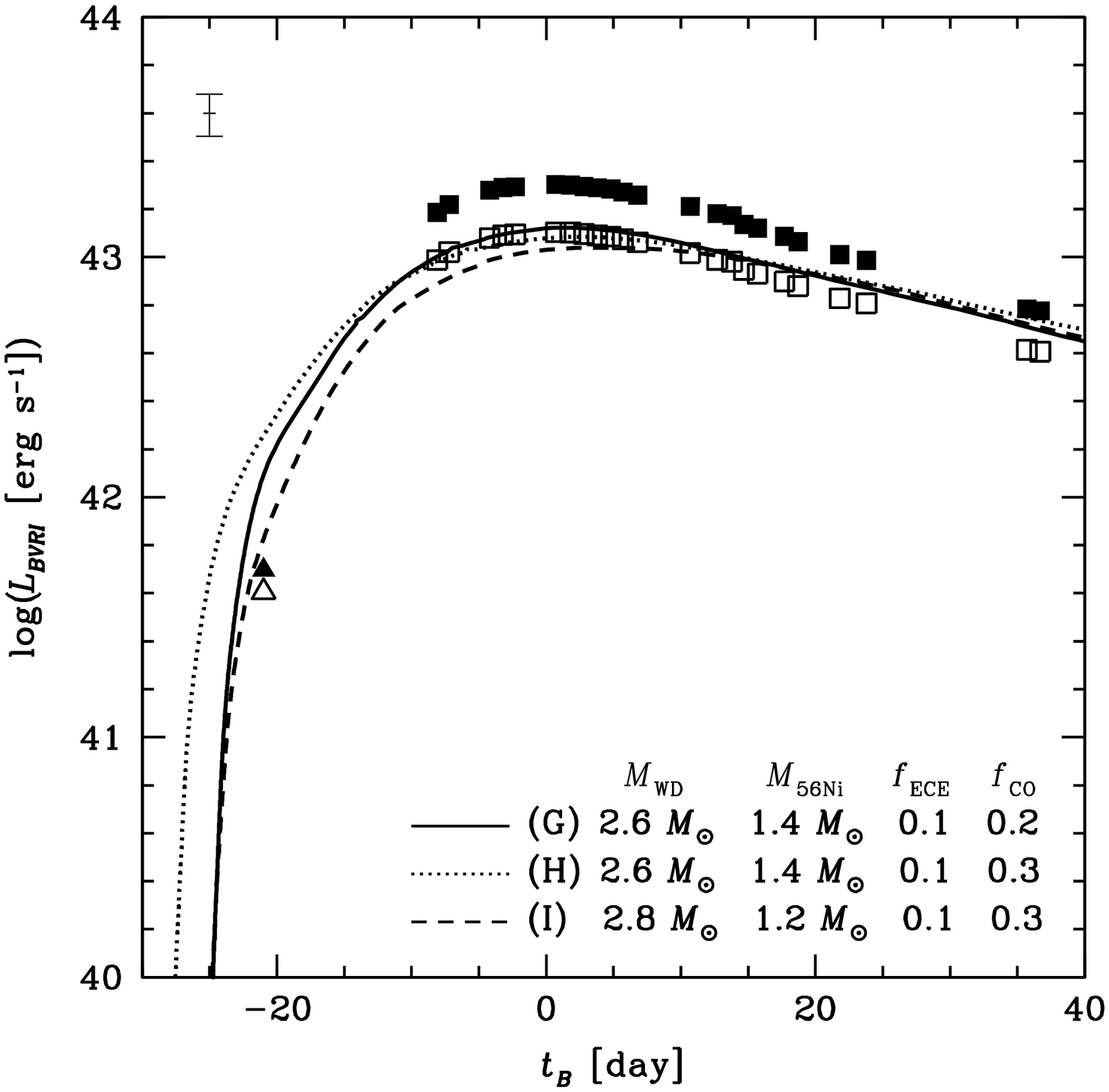}{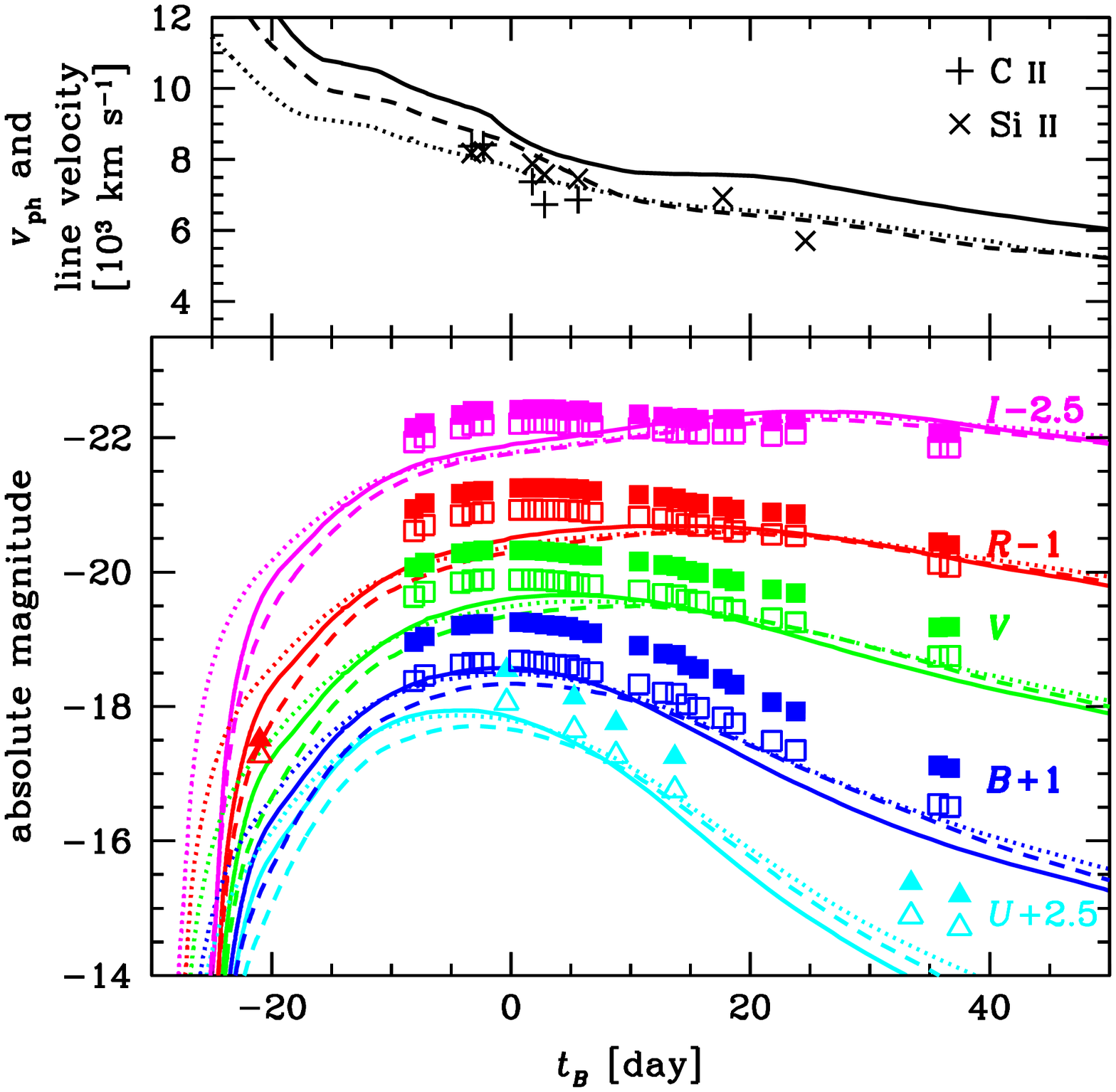}
\caption{Same as Figure \ref{best-fitted model}, but for models {\tt G}, {\tt H}, and {\tt I}.}
\end{figure*}

\begin{figure*}
\plottwo{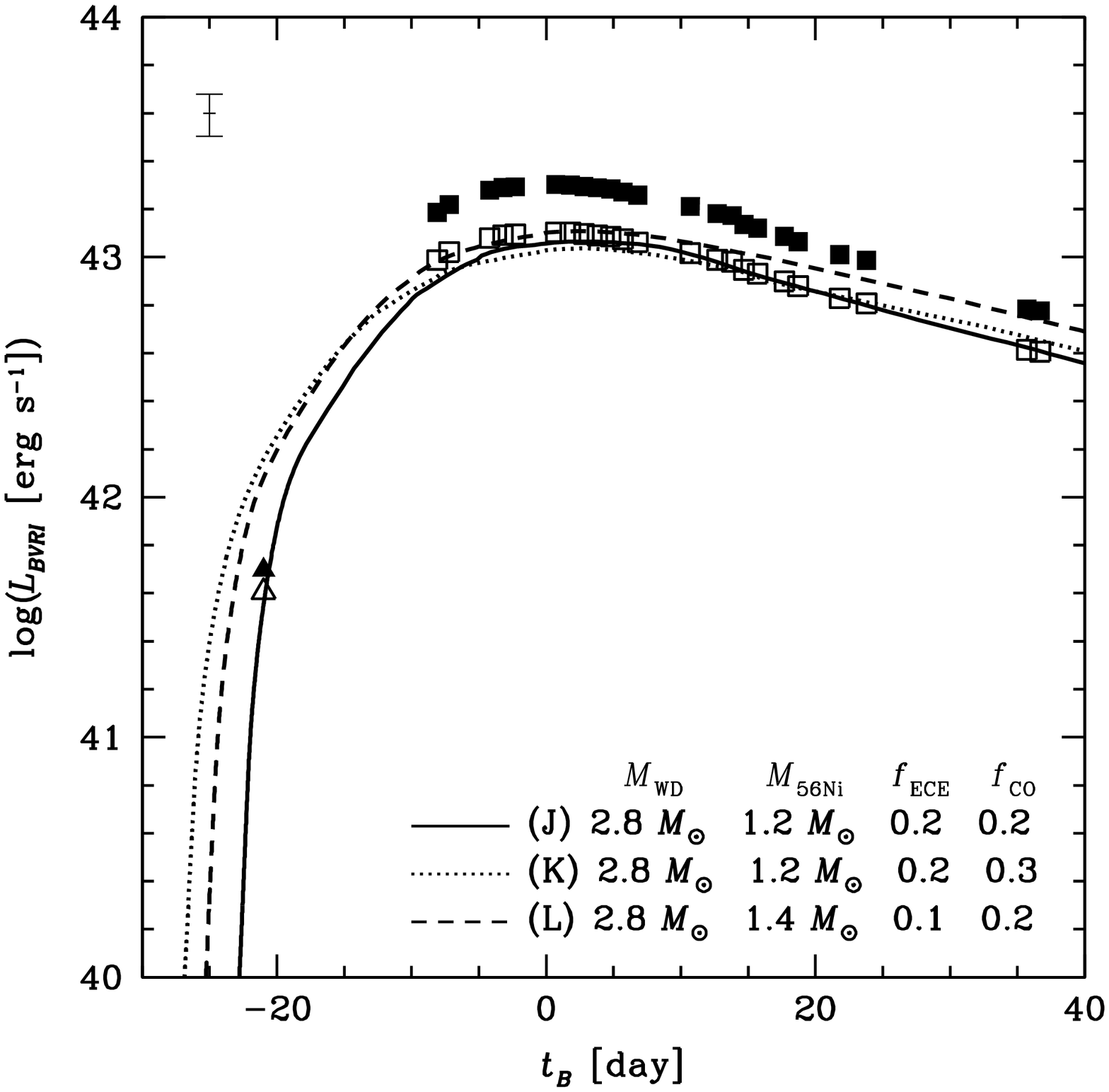}{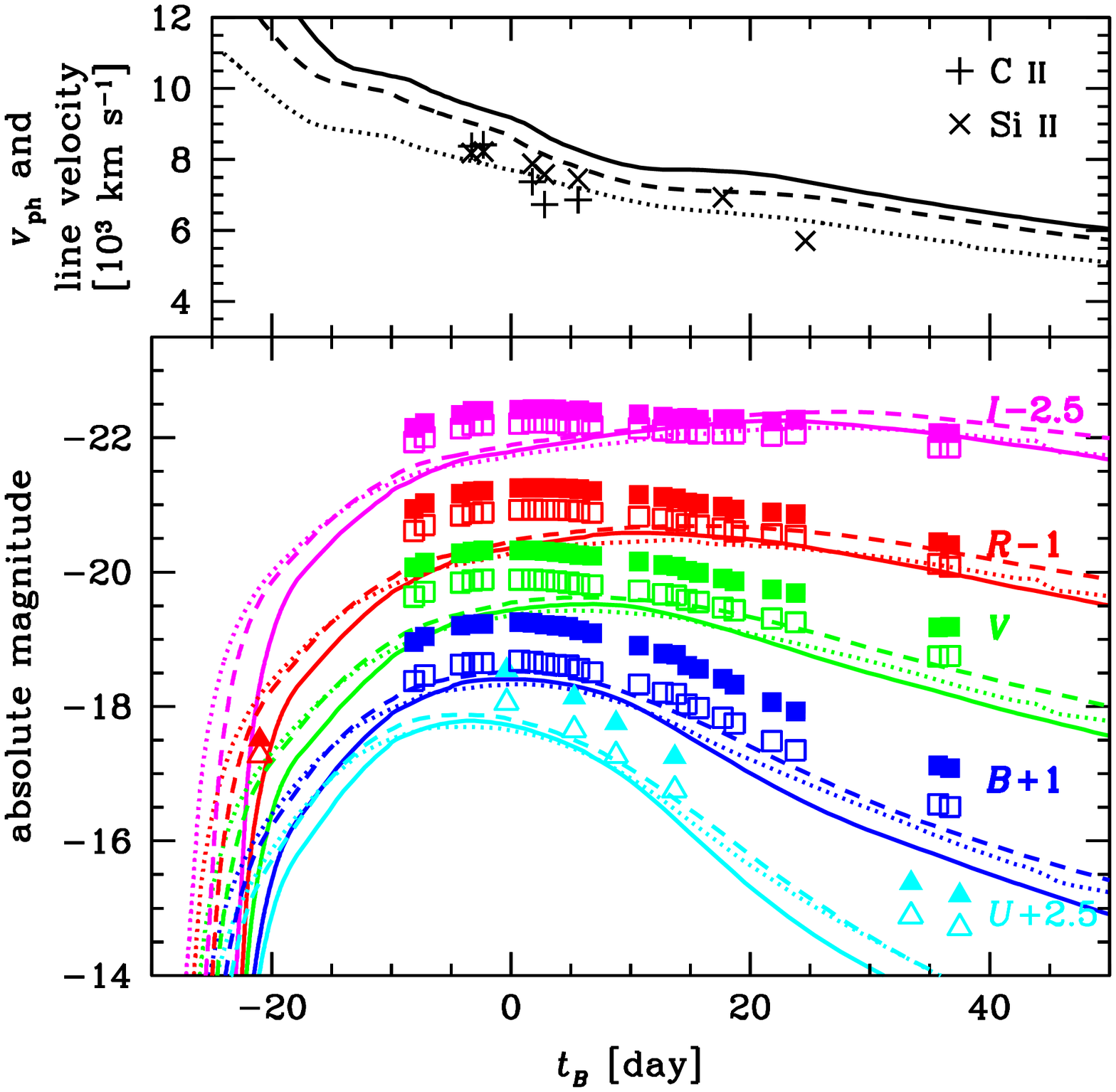}
\caption{Same as Figure \ref{best-fitted model}, but for models {\tt J}, {\tt K}, and {\tt L}.}
\end{figure*}

\begin{figure*}
\plottwo{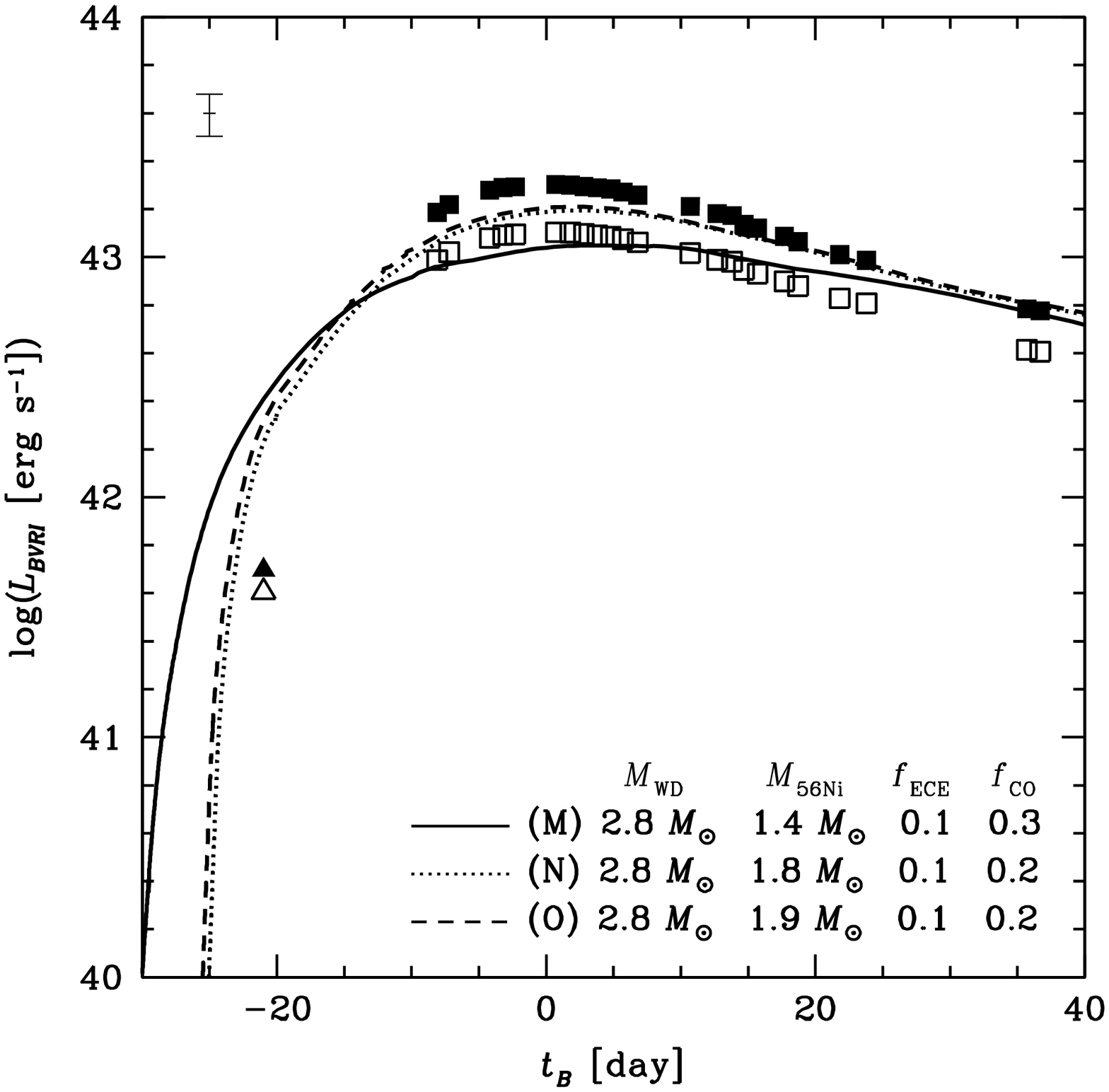}{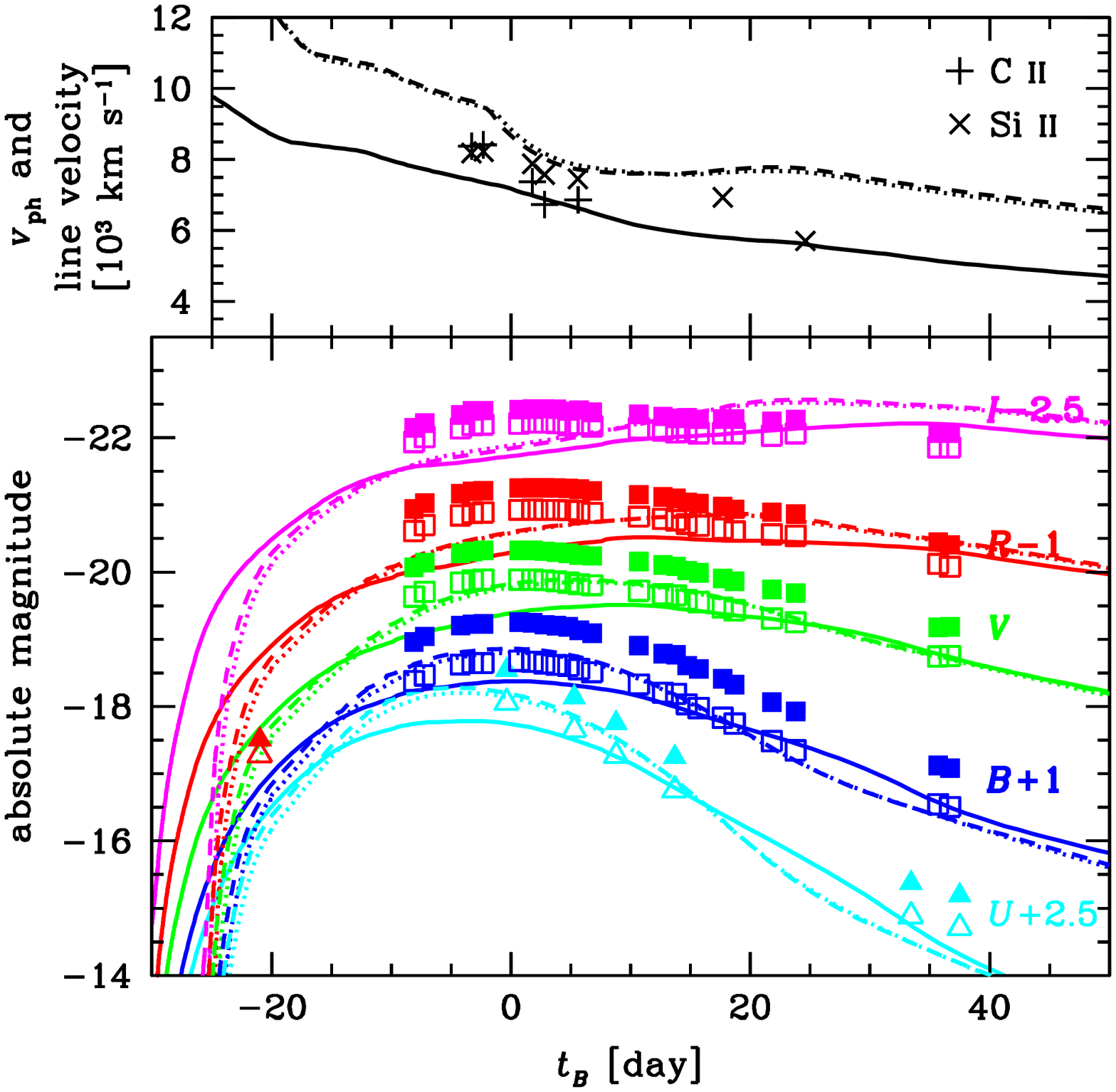}
\caption{Same as Figure \ref{best-fitted model}, but for models {\tt M}, {\tt N}, and {\tt O}.\label{best-fitted model2}}
\end{figure*}

With Figures \ref{best-fitted model}--\ref{best-fitted model2}, we make more detailed comparisons of models {\tt A}--{\tt O} with the observational data of SN 2009dc.
The comparisons are summarized as follows:

\begin{itemize}
\item The early detection in \citet{Silverman11} enables us to constrain the rising time of the models.
The $R$-band LCs of models {\tt E}, {\tt H}, and {\tt M} are too bright at $\tb\sim-20$ days regardless of whether the host-galaxy extinction is corrected or not.
These models can be excluded because their LCs evolve too slowly.

\item The remaining models for the case of the neglected host-galaxy extinction ({\tt A}--{\tt D}, {\tt F}, {\tt G}, and {\tt I}--{\tt L}) have as small $\Mni$ as 1.2 $\Ms$ and 1.4 $\Ms$, which is consistent with the analytical estimates.
While model {\tt J} has the smallest WRMSR, its $\vph$ is relatively larger than the observations.
Models {\tt D}, {\tt G}, and {\tt L} also have larger $\vph$ around $\tb\sim0$ day.
Among the other models ({\tt A}--{\tt C}, {\tt F}, {\tt I}, and {\tt K}), the models with $\Mwd\leq2.4$ $\Ms$ ({\tt A}--{\tt C}) are preferred, because their \bvri\ LCs reproduce the observations well around the peak.
Especially, model {\tt B} has the smallest $\vph$ of the three models, so we suggest that model {\tt B} is the most plausible model for SN 2009dc without the host-galaxy extinction.

\item If the extinction by the host galaxy is significant, we have only two candidate models with $\Mwd=2.8$ $\Ms$ ({\tt N} and {\tt O}).
They both have similar properties; slightly larger $\Lbvri$ at $\tb\sim-20$ days, but a bit smaller $\Lbvri$ around $\tb\sim0$ day.
We regard model {\tt N} as the most plausible model for the extinction-corrected case because the rising time is shorter than model {\tt O}\@.
\end{itemize}

By these results, we suggest that the mass of progenitor WD of SN 2009dc is 2.2--2.4 $\Ms$ if the extinction by its host galaxy is negligible, and $\sim$2.8 $\Ms$ with the extinction if the extinction is significant.
The $^{56}$Ni mass needed for the SN is 1.2--1.4 $\Ms$ for the former case, and 1.8--1.9 $\Ms$ for the latter, respectively.
For the latter case with extinction, the estimated masses are consistent with those suggested by \citet{Taubenberger11}, though they assume that the mean optical opacity of SN 2009dc is similar to (normal) SN 2003du.

For SN 2007if, one of the super-Ch candidates, \citet{Scalzo10} estimate the total mass of its progenitor to be $\sim$2.4 $\Ms$, as massive as we estimate for SN 2009dc.
They consider a shell-structured model to explain the low \ion{Si}{2} line velocity and its plateau-like evolution, where the massive envelope decelerates the outer layers of the ejecta.
Our calculations also reproduce the lower \ion{Si}{2} line velocity and similar evolution of SN 2009dc (the right-top panels of Figures \ref{best-fitted model}--\ref{best-fitted model2}), and suggest that the low line velocity and flat evolution of \ion{Si}{2} can be explained by scaled super-Ch-mass WD models, 
as well as the shell-shrouding models.

\pagebreak

\section{Discussion}
\label{Discussion}

\subsection{Light Curves of Super-Chandrasekhar-Mass White Dwarf Models}
\subsubsection{Multi-Color Light Curves}

We have calculated the multi-color LCs of the super-Ch-mass WD model
for the first time (right-bottom panels in Figures \ref{best-fitted model}--\ref{best-fitted model2}).
Even the model being in good agreement with $\Lbvri$ and $\vph$ shows discrepancy for each band to some extent, especially in the $I$ band.
This deviation from observed SNe Ia in the $I$ band is also seen for a Chandrasekhar-mass WD model calculated by the code STELLA \citep[][Figure 9]{Woosley07}.
It could be improved by taking more spectral lines into account for the opacity calculation.

The comparison with the velocity suggests that the mixing occurred in the ejecta.
It is interesting to note that LCs of SN 2009dc in the $I$ band do not clearly show double-peak features.
For Chandrasekhar-mass WD models, \citet{Kasen06} calculated the multi-color LCs, to conclude that mixing in the ejecta could produce the single-peak LCs in the $I$ band.
In fact, our models are partially mixed and their $I$-band LCs do not show two peaks (Figures \ref{best-fitted model}--\ref{best-fitted model2}).
It could be also the case for super-Ch-mass WD models that mixing affects their $I$-band LCs, although some uncertainties mentioned above should be taken into account.

\subsubsection{Bolometric, \uvoir, and \bvri\ Light Curves}

\begin{figure}
\plotone{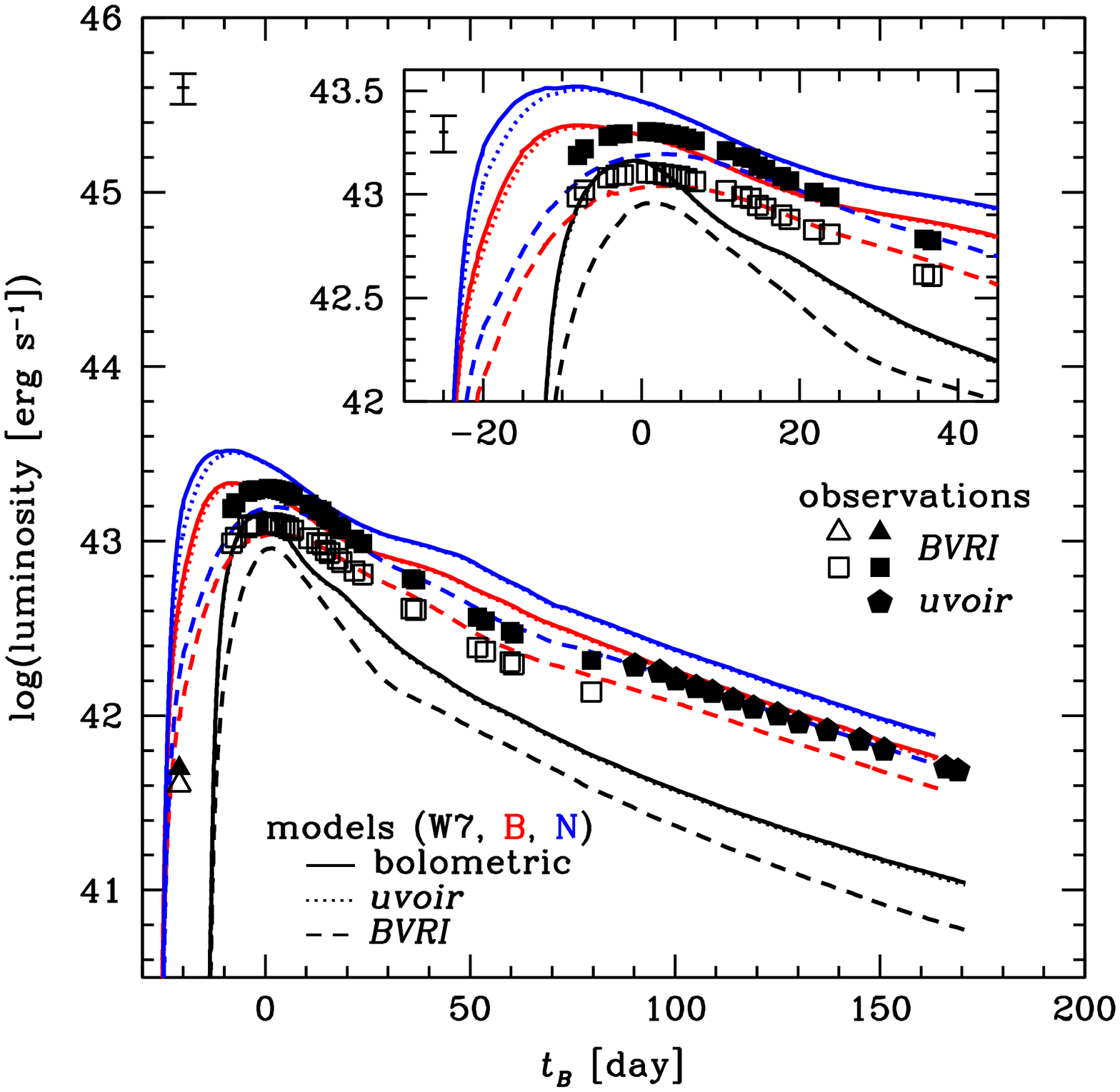}
\caption{Calculated bolometric (solid), \uvoir\ (dotted), and \bvri\ (dashed) LCs for models {\tt W7} (black), {\tt B} (red), and {\tt N} (blue).
The observed LCs are also shown as the open and filled symbols for the cases with and without the host-galaxy extinction, respectively.
For the open triangle and squares, the squares are same as in the right panel of Figure \ref{bolometric LCs}.
We assume $\Ehost=0.10$ mag for the filled triangles.
The pentagons are taken from Figure 7 of \citet{Taubenberger11}, where $\Emw=0.071$ mag and $\Ehost=0.10$ mag.\label{bolometric}}
\end{figure}

In Figure \ref{bolometric}, the bolometric, \uvoir, and \bvri\ LCs are plotted for models {\tt W7}, {\tt B}, and {\tt N}.
The \uvoir\ LCs peaks earlier than the \bvri\ LCs, by $\sim$2 days for {\tt W7} model, and by $\sim$10 days for model {\tt B} and {\tt N}, respectively.
As for the peak luminosity, the \uvoir\ LC of {\tt W7} is brighter 
than the \bvri\ LCs by $\sim$0.2 dex, while those of the two super-Ch-mass WD models, by $\sim$0.3 dex.
These shifts are understood by considering the ultraviolet (UV) radiation in the early phase \citep{Blinnikov00b}.

Most of \bvri\ LCs for the super-Ch-mass WD models in this paper seems to be fainter for $\tb<0$ day and slightly brighter after that, than the observations (Figures \ref{best-fitted model}--\ref{bolometric}).
This discrepancy could be solved by changing the $^{56}$Ni distribution in the model, which powers its radiation.
The more $^{56}$Ni outside could shorten the rising time, or make more luminous before the peak, while the less $^{56}$Ni inside could make fainter after that.
Further detailed modeling is needed since we simply assume the $^{56}$Ni distribution analogous to {\tt W7}.

In Figure \ref{bolometric}, we add the observation data for $\tb>90$ days plotted in Figure 7 of \citet{Taubenberger11} as filled pentagons, which are also corrected for the host-galaxy extinction by using their values.
The calculated bolometric LC fits to the observed LC tail quite well.
The observed tail lie just on model {\tt N}, which means that the estimate on $\Mni$ for the model is consistent. 

\subsection{Instabilities of Rotating White Dwarfs}

We calculate $\Eb$ by just extrapolating the formula, valid only for $\Mwd\gtrsim2.1$ $\Ms$ \citet{Yoon05}.
Using their Equations (22) and (29)--(31), we can calculate the ratio of the rotational energy ($T$) to (the magnitude of) gravitational one of ($W$) a super-Ch-mass WD with our parameter $\rhoc$.
This ratio $T/W$ indicates the stability of the WD.
If $T/W$ is small enough, the WD is stable against rotation.
If $T/W$ reaches 0.14, the WD suffers the non-axisymmetric instability \citep[e.g.][]{Ostriker73}.
The above equations show that a WD with $\Mwd\sim2.4$ $\Ms$ has $T/W=0.14$ for our $\rhoc$.
A super-Ch-mass WD with $\Mwd\geq2.4$ $\Ms$ might not form.
However, $T/W$ as a function of $\Mwd$ differs with the rotation law \citep[e.g.,][]{Hachisu86}, so that it is possible that even such a massive WD as $\Mwd\geq2.4$ $\Ms$ has $T/W < 0.14$ and thus form.
We thus consider the models with $\Mwd\geq2.4\Ms$.

Our results in this paper suggest that the progenitor WD mass of SN 2009dc exceeds 2.2 $\Ms$ even for the case where the host-galaxy extinction is negligible.
Such a large $\Mwd$ might suggest that the explosion of this super-Ch-mass WD is triggered when its mass reaches the critical mass for the above instability of the rotating WDs \citep{Hachisu12}.

\subsection{Possible Progenitors and Scenarios}

The progenitor WD mass for SN 2009dc in our estimate largely exceeds $\Mch$, thus putting severe constraints on the presupernova evolution of the binary system.
This also should have important implications on the progenitor scenarios of ordinary SNe Ia, both the single degenerate (SD) and double degenerate (DD) scenarios.

\subsubsection{Single Degenerate Scenario}

The initial masses of the WD ($\Mwdi$) and its companion star ($\Mcompi$) should be sufficiently large at the beginning of accretion in order to increase $\Mwd$ to $\gtrsim$2.4 $\Ms$.
\citet{Chen09} argued that even for $\Mwdi=1.24$ $\Ms$ the accretion from the companion of $\Mcompi\lesssim3.5$ $\Ms$ does not obtain $\Mwd\gtrsim1.7$ $\Ms$.
However, they does not take into account the effect of mass-stripping from the companion star due to the strong WD wind \citep{Hachisu07}.
Because the mass-stripping effectively reduces the mass transfer rate from the companion to the WD, the companion star can be as massive as $\Mcompi\sim4$--7 $\Ms$ and $\Mwd$ can reach $\sim$2.4 $\Ms$ \citep{Hachisu12}.
Still, to realize such a massive WD, $\Mwdi\gtrsim1.1$ $\Ms$ is preferable \citep{Chen09,Hachisu12}.

The formation of the C+O WD with $\Mwdi\gtrsim1.1$ $\Ms$ is realized in the special binaries.  In stars of main-sequence mass of $\lesssim$8 $\Ms$, the C+O core mass is $<$1.07 $\Ms$ to avoid off-center C-ignition before the AGB phase \citep[e.g.,][]{Umeda99}.
After the dredge-up of the He layer, the C+O core increases its mass during the AGB phase, if the binary separation is wide enough to accommodate the AGB star.
In such a binary system, the C+O WD with $\Mwdi\gtrsim1.1$ $\Ms$ can be formed, if the C+O core of the AGB star has already grown massive when the AGB envelope is lost in a wind or by Roche-lobe overflow.
Thus $\Mwdi$ is larger if the mass loss rate from the AGB star is smaller.
Therefore, the C+O WDs with larger $\Mwdi$ is more likely to form in the lower metallicity system and in the initially wider binary \citep{Hachisu07}.
Since the binary must also be close enough for the mass-transfer to occur, the suitable binary system could be rare, which is consistent with the low occurrence frequency of the super-Ch-mass explosion.

These requirements of large enough $\Mwdi$ and $\Mcompi$ in the SD scenario predicts that SNe Ia from super-Ch-mass WDs are associated with the star-forming region and low metallicity
environment.
It is interesting to note that the hosts of the observed super-Ch candidates are faint and star-forming galaxies except for SN 2009dc \citep[e.g.][Table 7]{Taubenberger11}.
For SN 2009dc, the host galaxy UGC 10064 is a passive (S0) galaxy.
However, at $\sim$40 kpc away from UGC 10064, there is a blue irregular galaxy UGC 10063, which could also have a star formation in the recent past by the interaction \citep{Silverman11}.

\subsubsection{Double Degenerate Scenario}

For the DD scenario, to form a WD of $\gtrsim$2.4 $\Ms$ by merging of two C+O WDs, the primary C+O WD needs to be initially as massive as $\Mwdi\gtrsim1.33$ $\Ms$ because of the following reason.
To form a massive C+O core in the primary AGB star, the initial binary system needs to be wide enough.
The Roche lobe overflow of the primary AGB star is so rapid that a formation of a common envelope is unavoidable.
After the loss of mass and angular momentum from the common envelope, a primary C+O WD and the secondary star are left in a binary with a small separation.
If the separation is too small for the secondary star to become an AGB star, the mass of the secondary C+O WD is $\Mwdi\lesssim1.07$ $\Ms$.
In order to form a WD of $\gtrsim$2.4 $\Ms$, the primary C+O WD should be more massive than 1.33 $\Ms$, which would be very rare.
In the above scenario, formation of the double C+O WDs whose initial masses are both $\sim$1.2 $\Ms$ may not be possible because of the shrink of the binary system after the first common envelope phase.

\section{Conclusions}
\label{Conclusions}
To constrain the properties of SN 2009dc, we have calculated multi-band LCs for the exploding super-Ch-mass WD models with a range of model parameters.
We find that the mass of the WD and other model parameters are constrained as follows.

\begin{itemize}
\item The observed \bvri\ LCs of SN 2009dc are well-explained by the super-Ch-mass WD models with $\Mwd=1.8$--2.8 $\Ms$ and $\Mni=1.2$--1.8 $\Ms$, if the extinction by the host galaxy is negligible.
\item The observed line velocity of \ion{Si}{2} is consistent with $\vph$ of several models with $\Mwd=2.2$--2.8 $\Ms$, but significantly lower than $\vph$ of the less massive models.
\item Among our models, the most plausible model is model {\tt B} with $\Mwd=2.4$ $\Ms$ (1.2 $\Ms$ of $^{56}$Ni, 0.24 $\Ms$ of ECEs, 0.24 $\Ms$ of IMEs, and 0.72 $\Ms$ of C+O) for the case without the host-galaxy extinction.
\item If the extinction is considered, the mass of the super-Ch-mass WD needs to be as massive as $\Mwd\sim2.8$ $\Ms$ (i.e. model {\tt N} with 1.8 $\Ms$ of $^{56}$Ni, 0.28 $\Ms$ of ECEs, 0.16 $\Ms$ of IMEs, and 0.56 $\Ms$ of C+O).
We find that the fit to the observation is less successful for model {\tt N} than model {\tt B}\@.
\item Such a large $\Mwd$ might suggest that the explosion of the super-Ch-WD might be related to the onset of the instability of the differentially rotating WD.
\end{itemize}

Our results in this paper are based on the simplified models of the super-Ch-mass WDs.
There are still several uncertainties in the models and LCs; such as the parameterization of the models, the opacity calculated by the code, aspherical effects, and effects of possible circumstellar interaction, as well as the instability of the massive WDs.
However, $\Mwd$ and $\Mni$ of the plausible models for SN 2009dc are quite consistent with the observations, suggesting that our present approach works well for this super-Ch candidate.

\acknowledgments

We are grateful to Masayuki Yamanaka for providing us the detailed observation data of SN 2009dc, to Keiichi Maeda and Nozomu Tominaga for the constructive discussion on the construction and LC calculations of super-Ch-mass WDs models.
Y.K. acknowledges the Japan Society for the Promotion of Science (JSPS) for support through JSPS Research Fellowships for Young Scientists, and his work is supported by Grant-in-Aid for JSPS Fellows \#22$\cdot$7641.
The work of S.I.B. and E.I.S. in Japan is supported by the Ministry of Education, Culture, Sports, Science and Technology; and in Russia by grants RFBR 10-02-00249-a and 10-02-01398-a, the Grant of the Government of the Russian Federation (No.~11.G34.31.0047), Sci.~Schools-3458.2010.2 and -3899.2010.2, and a grant IZ73Z0-128180/1 of the Swiss National Science Foundation (SCOPES).
This research has been supported in part by the Grant-in-Aid for Scientific Research of MEXT (22012003, 22840009, and 23105705) and JSPS (23540262) and by World Premier International Research Center Initiative, MEXT, Japan.

\end{document}